%% file: Optica-template.tex
\documentclass{optica-article}
\usepackage{subcaption}
\journal{opticajournal} 

\articletype{Research Article}

\usepackage{xcolor}

\newcommand{\ve}[1]{\mathbf{#1}}
\newcommand{\te}[1]{\overline{\overline{#1}}}

\usepackage{lineno}

\begin{document}

\title{Step function in momentum space by a metagrating}

\author{Mahmoud A. A. Abouelatta\authormark{1,*}, Sergejs Boroviks\authormark{2}, Olivier J. F. Martin\authormark{2}, and Karim Achouri\authormark{1}}

\address{\authormark{1}Laboratory for Advanced Electromagnetics and Photonics, École Polytechnique Fédérale de Lausanne (EPFL), 1015 Lausanne,
Switzerland
\\
\authormark{2}Nanophotonics and Metrology Laboratory, École Polytechnique Fédérale de Lausanne (EPFL), 1015 Lausanne,
Switzerland\\
}

\email{\authormark{*}mahmoud.abouelatta@epfl.ch} 


\begin{abstract*} 
Metasurface research has shown significant potential for controlling the polarization, amplitude, phase and propagation direction of light. Nevertheless, control over the 
angular response of incident light still remains a long-standing problem. In this work, we show the potential of diffractive systems for obtaining a step function in momentum space where the mirror symmetry of the angular transmittance is broken. By engineering the scattering response of an asymmetric particles in a metagrating, we could obtain such a step function in a passive, reciprocal, and lossless fashion. More specifically, 
the metagrating performs filtering in the momentum space with an abrupt switching from reflection to transmission for an incident electromagnetic wave with an arbitrary spatial profile. 
This metagrating may find diverse applications in the context of optical spatial analog computing. 
Moreover, it paves the way for exploring the capabilities of diffractive systems for gaining full control over the angular response of light using arbitrary momentum transfer functions.
\end{abstract*}

\section{Introduction}
Metasurfaces have enabled versatile ways to control the polarization, the temporal frequency response, the amplitude and the phase profiles of light. By engineering their electromagnetic response at the sub-wavelength scale, a plethora of metasurface concepts and applications have been proposed to manipulate these degrees of freedom~\cite{planar,metarev}. However, despite the great light propagation engineering capabilities that metasurfaces enable, the possible options they offer for controlling the spatial frequency of light (i.e., to control the angular response) remain limited. Hence, most metasurface applications have been limited to normal incidence or oblique propagation within the paraxial limit~\cite{karim_paraxial,karim_multipolar}. Achieving scattering control over the full angular spectrum would make spatial analog computing possible in ultra-compact systems where signals are processed at the speed of light, and are analyzed in the spatial Fourier domain~\cite{Fourier_karim}. In this area of research, several works have already shown the potential of metasurfaces for diverse applications such as spatial differentiation~\cite{differentiation}, planar retro-reflection~\cite{retroreflector}, refraction control~\cite{refraction}, and angular selective response~\cite{selectivity}, among many others~\cite{review_analog,review_romain}. One of the milestones in the research associated with the angular response of light are the generalized laws of refraction and reflection~\cite{generalized}, which model how light may be deflected by the introduction of planar phase gradients. However, a careful examination of these laws showed that they are merely an approximation to the diffraction theory~\cite{grating_tutorial}, implication being that they do not provide rigorous solutions to Maxwell's equations. Consequently, they can not, for example, describe quantitatively the total efficiency of beam steering in a system~\cite{angular_modelling} and do not allow for fully efficient refraction operations~\cite{refraction_karim}. This suggests that a more rigorous formalism should be considered to achieve an advanced control of the angular response of light.

Among the various problems associated with the angular light scattering control,  breaking the symmetry of the angular transmission through an optical system is especially interesting in the context of this work. Several examples of asymmetric metasurfaces that break the angular symmetry of the transmittance have already been proposed. For instance, by sandwiching a metasurface between two polarizers and manipulating the incident polarization state, the angular symmetry of the optical transfer function may be broken~\cite{asymm_OTF}. Few other works have reported angularly asymmetric absorbance (A) by relying on the anomalous Brewster effect~\cite{brewster_asymm}, and evanescent field engineering via a phase-gradient (i.e., spatially-varying) metasurface~\cite{evancent_eng}. It is important to note that all of these designs require, in one way or another, asymmetric angular losses in order to achieve asymmetric angular transmittance. This begs the question as to whether or not it is possible to achieve asymmetric angular transmittance in a lossless fashion. It turns out that, as we show thereafter, it is in fact impossible to achieve lossless asymmetric angular transmittance with a planar non-diffractive reciprocal optical system.

To overcome this limitation, we shall therefore explore the potential of diffractive surfaces, e.g. metagratings, to achieve such an optical response. In contrast to metasurfaces, metagratings have extra degrees of freedom for manipulating electromagnetic waves in multiple ports. As such, several metagrating schemes have been introduced in the literature for extreme beam steering~\cite{metagrating_alu}, parallel
computing~\cite{fleury}, solving integral equations~\cite{integral}, and spatiotemporal pulse shaping~\cite{spatiotemp,spatiotemp2, spatiotemporal_exp}.

In this work, we demonstrate how asymmetric angular transmittance may be theoretically achieved with a lossless reciprocal diffraction grating. We then proceed to fabricate such a device and experimentally verify its asymmetric scattering properties. We anticipate that asymmetric transmittance may find applications in optical analog signal processing devices, for instance, to achieve image filtering in an ultra-thin miniaturized fashion. A typical example would be that of a Schlieren imaging system, which relies on the concept of filtering out a part of the momentum space with a knife edge, that may be implemented without a bulky free-space system composed of multiple lenses or mirrors. Moreover, it could also span further applications such as angular dependent smart windows where the Sun energy needs to be reflected for a desired angular range (instead of being absorbed by the system)~\cite{thermal}.

\section{Theoretical considerations}

\begin{figure}
\centering
\includegraphics[width=\textwidth]{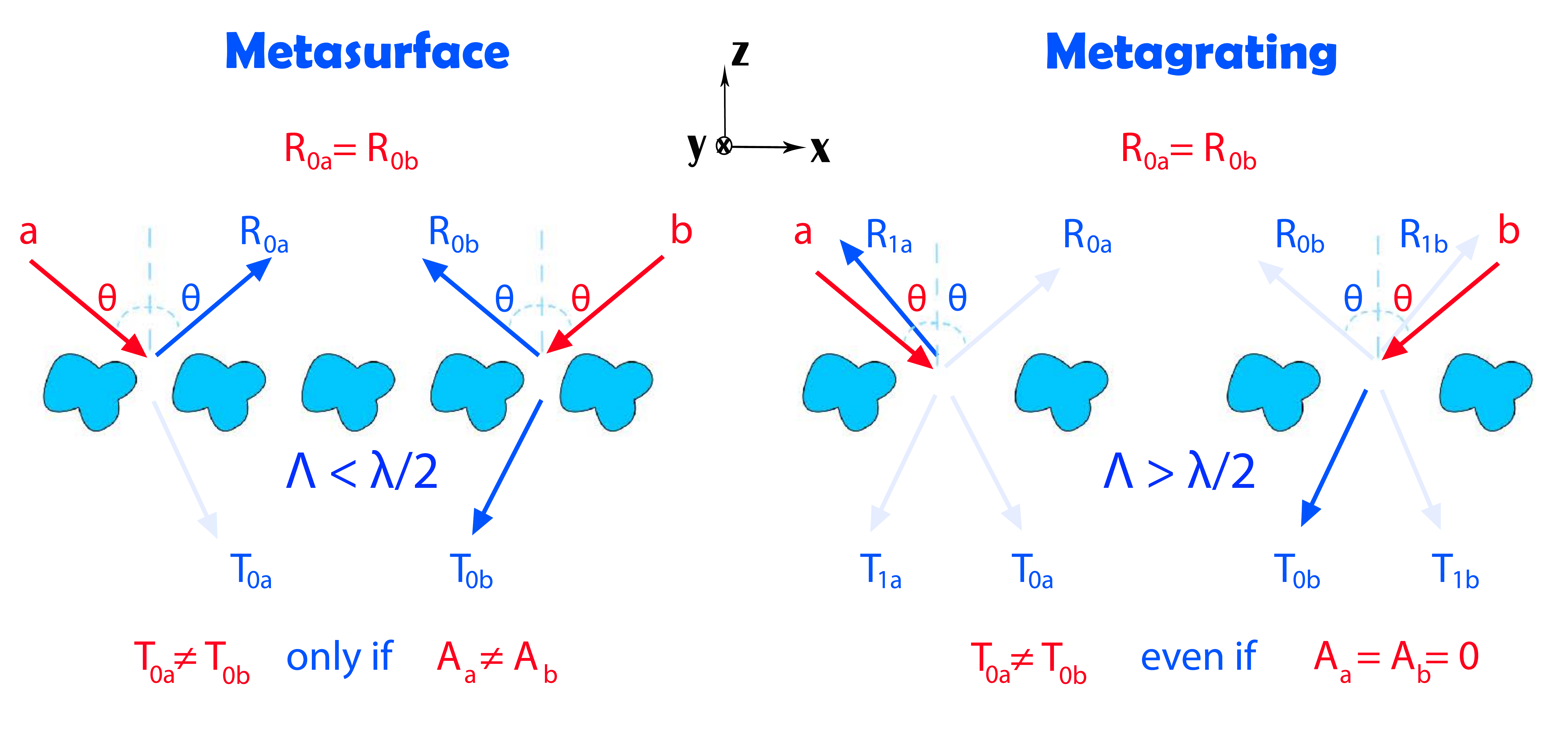}
\caption{Illustration of the key differences between a non-diffractive (metasurface) and a diffractive array of scatterers (metagrating). Left side: the angular symmetry of the $0^\text{th}$-order reflectance is enforced by reciprocity, whereas the angular symmetry of the transmittance can be broken by introducing asymmetric absorbance ($A_{a}\neq A_{b}$). Right side: the angular symmetry of the transmittance is broken by allowing the system to be diffractive while engineering the $0^\text{th}$-order reflectance to have a value close to zero. $R_{0a}$, $R_{1a}$, $T_{0a}$ and $T_{1a}$ denote the $0^\text{th}$-order, first-order reflectance, $0^\text{th}$-order, and first-order transmittance, respectively, when a plane wave is incident from port ``a'', and similarly for waves incident from port ``b''. $A_{a}$ and $A_{b}$ represent the absorbance for incidence from port ``a'' and ``b'', respectively, while $\Lambda$ is the array period.
}
\label{fig:reciprocity}
\end{figure}

The underlying principle of how angular asymmetry has been previously achieved using non-diffractive metasurface platforms is illustrated in the left side of Fig.~\ref{fig:reciprocity}. Owing to the fact that the reciprocity of such systems enforces the reflectance (R) to have angular symmetry (i.e., $R(\theta)=R(-\theta)$), the \emph{only} possible way to break the symmetry of the angular transmittance (i.e., $T(\theta)\neq T(-\theta)$) is by having angularly asymmetric absorption~\cite{lukas_asymm}. This is straightforward to demonstrate, by considering that
\begin{equation}
T(\theta) = 1 - R(\theta) - A(\theta) \quad \text{and} \quad T(-\theta) = 1 - R(-\theta) - A(-\theta),
\end{equation}
which implies that $T(\theta)\neq T(-\theta)$ is only possible if $A(\theta)\neq A(-\theta)$.

The concept that we introduce in this work is shown in the right side of Fig.~\ref{fig:reciprocity}. By allowing propagation in multiple diffraction directions, and by appropriately breaking the spatial symmetry of the system, the angular transmittance may have an asymmetric response \emph{without} enforcing the system to be dissipative. Additionally, to maximize this angular transmission asymmetry, and thus maximize the efficiency of the system, it is necessary to minimize the $0^\text{th}$-order reflectance, which is angularly symmetric by reciprocity, as explained above.

\begin{figure}
\centering
\includegraphics[width=\textwidth]{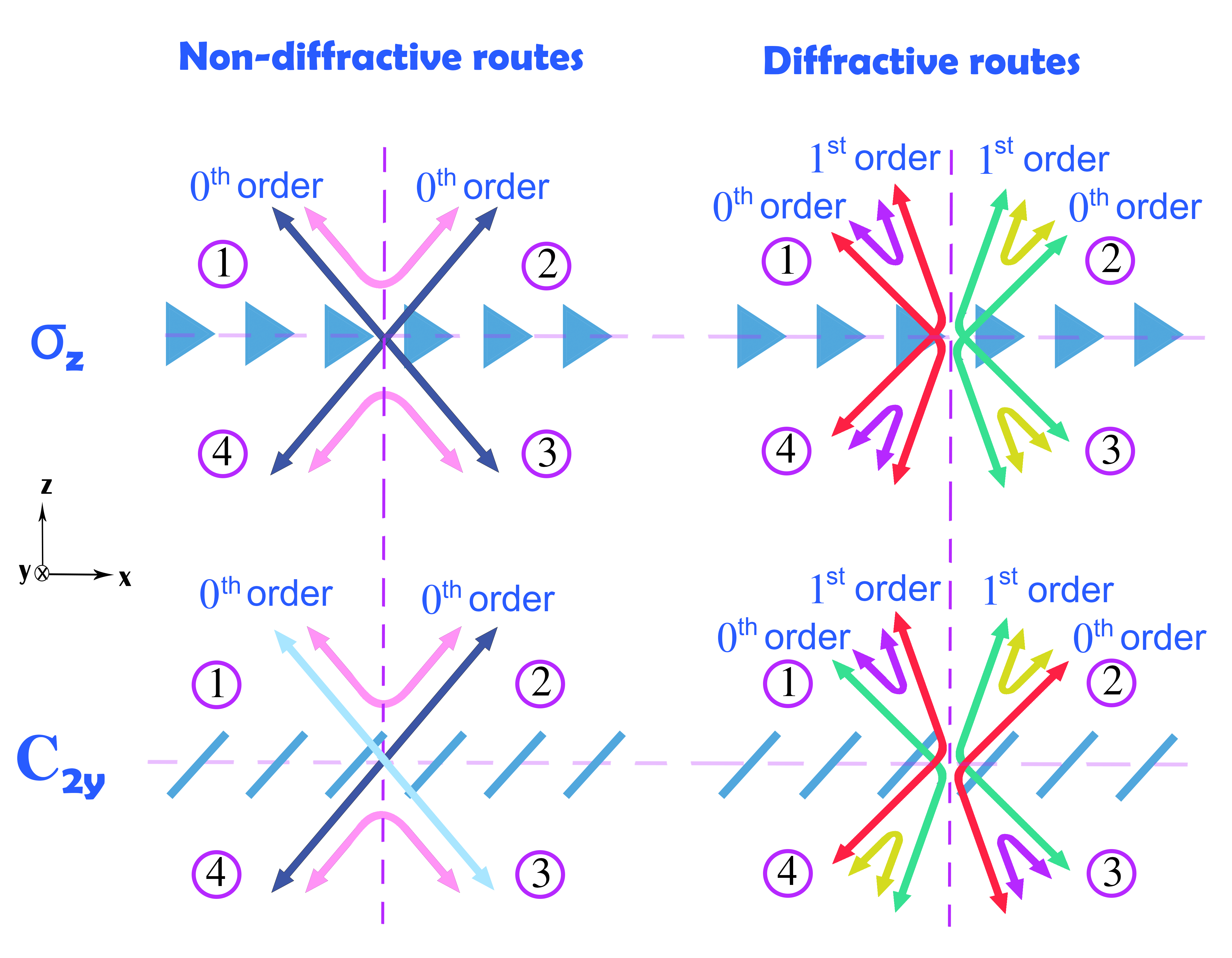}
\caption{Non-diffractive and diffractive channels of wave propagation in systems with two different kinds of spatial symmetries. The top row corresponds to mirror symmetry in the direction of the $z$-axis ($\sigma_{z}$). The bottom row corresponds to rotation symmetry by $180^{\circ}$ ($\rm C_{2y}$) around the $y$-axis. The arrows in the figure represent the scattering between the different ports of the various quadrants and identical colors indicate identical scattering parameters. The arrows are double sided to visualize the reciprocity of the system in different scattering channels.
}
\label{fig:symmetry}
\end{figure}

Thus far, we discussed the metagrating concept in the framework of the reciprocity theorem. We shall now illustrate it from the perspective of spatial symmetries. It is insightful to consider which kinds of symmetries may be preserved, and which ones should be broken, in order to achieve angular asymmetry around the $z$-axis, as illustrated in the right side of Fig.~\ref{fig:reciprocity}. Such insights allow us to find the simplest possible structure for this purpose and provide a powerful approach to optimize the scattering response of the system. For example, a unit cell which has mirror symmetry ($\sigma_{x}$) and periodicity along the $x$-axis can not have angular asymmetry around the normal direction (i.e., the $z$-axis). However, when it comes to mirror symmetry along the $z$-axis ($\sigma_{z}$), the implications are more involved. For this reason, we have developed a formalism that allows us to easily and rigorously deduce the scattering symmetries in a diffraction grating (with one-dimensional periodicity) based on the spatial symmetries of its unit cell. This formalism is described in details in the Supplementary Section 1. For an array that is infinite along the $y$-direction, the scattering symmetries corresponding to two different possible unit cell spatial symmetries, i.e., only possessing $\sigma_z$ or $C_{2y}$, are illustrated in Fig.~\ref{fig:symmetry}. The formulas for the scattering matrices corresponding to $\sigma_z$ and $C_{2y}$ symmetries are shown in Eqs. S9 and S10, respectively. We consider that the incident wave is impinging on the array in the plane of symmetry ($xz$-plane) from quadrants denoted 1 to 4 at an arbitrary angle. On the left side of the figure, we show the scattering symmetries between the non-diffractive channels ($0^\text{th}$ order) whereas the right side of the figure shows the coupling between the $0^\text{th}$ order channels and the $1^\text{st}$ order diffractive channels. 

As can be seen in the top-left of Fig.~\ref{fig:symmetry}, when a wave is incident from quadrant 1 to 3 or from quadrant 4 to 2, the $0^{\rm th}$ order transmittance has to be the same due to the mirror symmetry along the $z$-axis ($\sigma_{z}$). Reciprocity also enforces the transmittance from quadrant 4 to 2 and from quadrant 2 to 4 to be the same. Thus, considering these two results, the system has to be angularly symmetric for the $0^{\rm th}$ order transmittance demonstrating that a structure with only $\sigma_z$ symmetry (and broken $\sigma_x$ symmetry) \emph{cannot} be used for angularly asymmetric transmittance in the $0^{\rm th}$ order. Nevertheless, rerouting the energy into the $1^{\rm st}$ order transmittance or reflectance may be angularly asymmetric, as shown in the upper right side of Fig.~\ref{fig:symmetry}. 

In the case where the structure exhibits only a $C_{2y}$ symmetry,  symmetry breaking in both the $0^{\rm th}$ and $1^{\rm st}$ order transmittance channels is possible, as shown in the bottom-left of Fig.~\ref{fig:symmetry}. Additionally, the $1^{\rm st}$ order reflectance is also asymmetric,implying that a structure possessing $C_{2y}$ symmetry is the simplest possible structure to achieve our desired asymmetric scattering response\footnote{It may also be achieved with a structure with fully broken symmetry at the cost of a more complicated fabrication procedure.}.

Besides reciprocity and symmetry perspectives, the metagrating could also be analyzed based on multipolar theory. We show the multipolar decomposition for a unit cell of the metagrating in Fig. S6, which proves that our system could be modeled as an array of two-dimensional tilted electric dipoles in the plane of symmetry (i.e., the $xz$-plane). Since the metagrating is composed of dipoles only, the nonlocal interactions between different unit cells is negligible, and consequently, the introduced metagrating's unit cell is local despite its strong angular dependence. In this context, a perspective article classifies metasurfaces into three categories (local nondiffractive metasurface, nonlocal nondiffractive metasurface, and nonlocal diffractive metasurface)~\cite{nonlocaldiff}. In contrast to these three categories, our metgrating could be regarded as a diffractive metasurface with unit cell harboring local field interactions. Thus demonstrating that nonlocal interactions are not a prerequisite for achieving significant angular dispersion. In a future work, we will show how such kind of systems composed of tilted dipoles could be analytically modeled using Floquet expansion and the polarizability tensor.

Based on the consideration of reciprocity and spatial symmetries, we come to the conclusion that, in order to achieve lossless reciprocal angular transmittance, we should implement a diffraction grating with $C_{2y}$ symmetry.

\section{Numerical and Experimental Results}

\begin{figure}
\centering
\includegraphics[width=\textwidth]{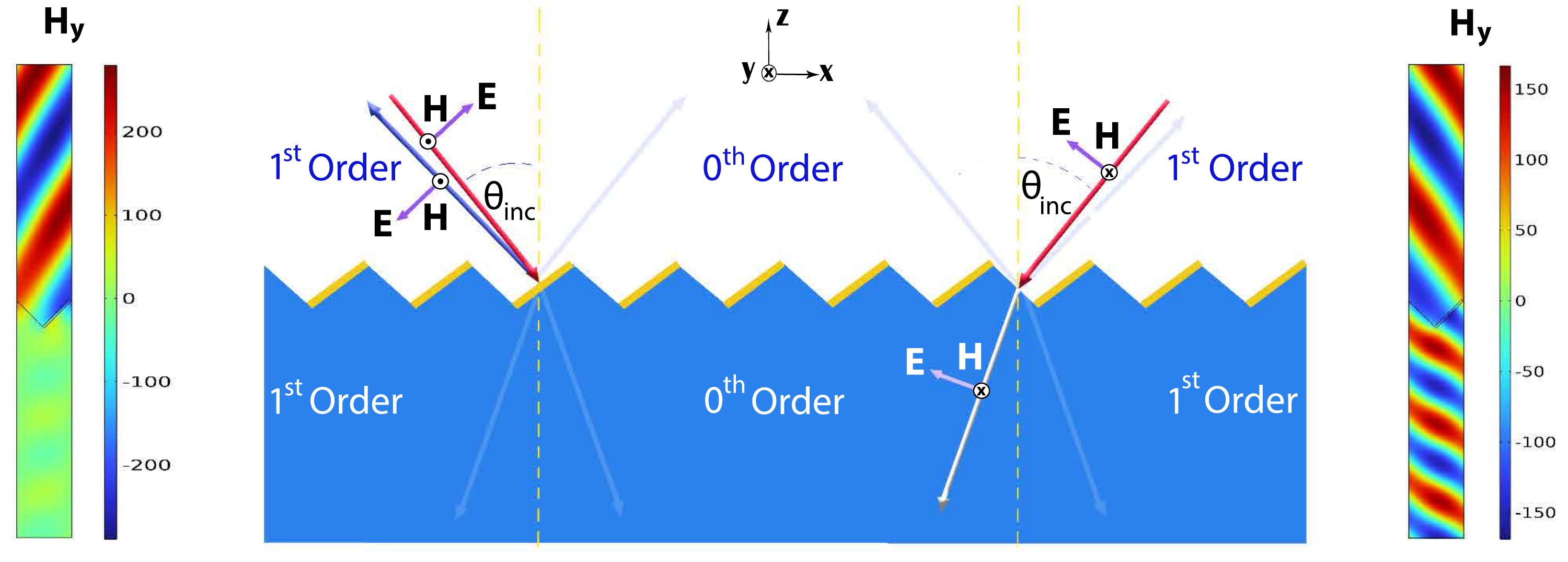}
\caption{Metagrating with angular asymmetric transmittance. Two cases are depicted where two incident TM polarized plane waves are retroreflected and transmitted for positive and negative incidence angles, respectively. The metagrating consists of an array of silica ridges  (shown in blue color) with asymmetrically deposited gold strips (shown in yellow color). The magnetic field profile ($ H_{y}$) is shown in the left and right sides for two TM plane waves incident on the metagrating with angles equal to $55^{\circ}$ and $-55^{\circ}$, respectively. The wavelength of the incident TM waves for these two field profiles is equal to 900 nm. The period of the grating ($\Lambda$) is 525 nm, while the triangular ridges are tilted with respect to the $x$-axis with an angle equal to $41^{\circ}$. The thickness of the gold strips is 22 nm. 
}
\label{fig:sch}
\end{figure}

The schematic of the proposed metagrating is depicted in Fig.~\ref{fig:sch}. It consists of triangular ridges made of silica with a thin layer of gold deposited on only one side of these ridges. The metagrating fully transmits transverse magnetic (TM) plane waves having negative angles of incidence with respect to the $x$-axis (i.e, from the right), as demonstrated by the magnetic field ($H_{y}$) distribution on the right side of Fig.~\ref{fig:sch}. However, the metagrating retro-reflects the incident TM plane waves which possess positive angles of incidence (i.e., from the left) where the energy is being rerouted into the first reflection diffraction order, as depicted in the magnetic field ($H_{y}$) distribution on the left side of Fig.~\ref{fig:sch}.

\begin{figure}[!htb]
\centering
\begin{subfigure}[b]{0.3\textwidth}
         \centering
         \includegraphics[width=\textwidth]{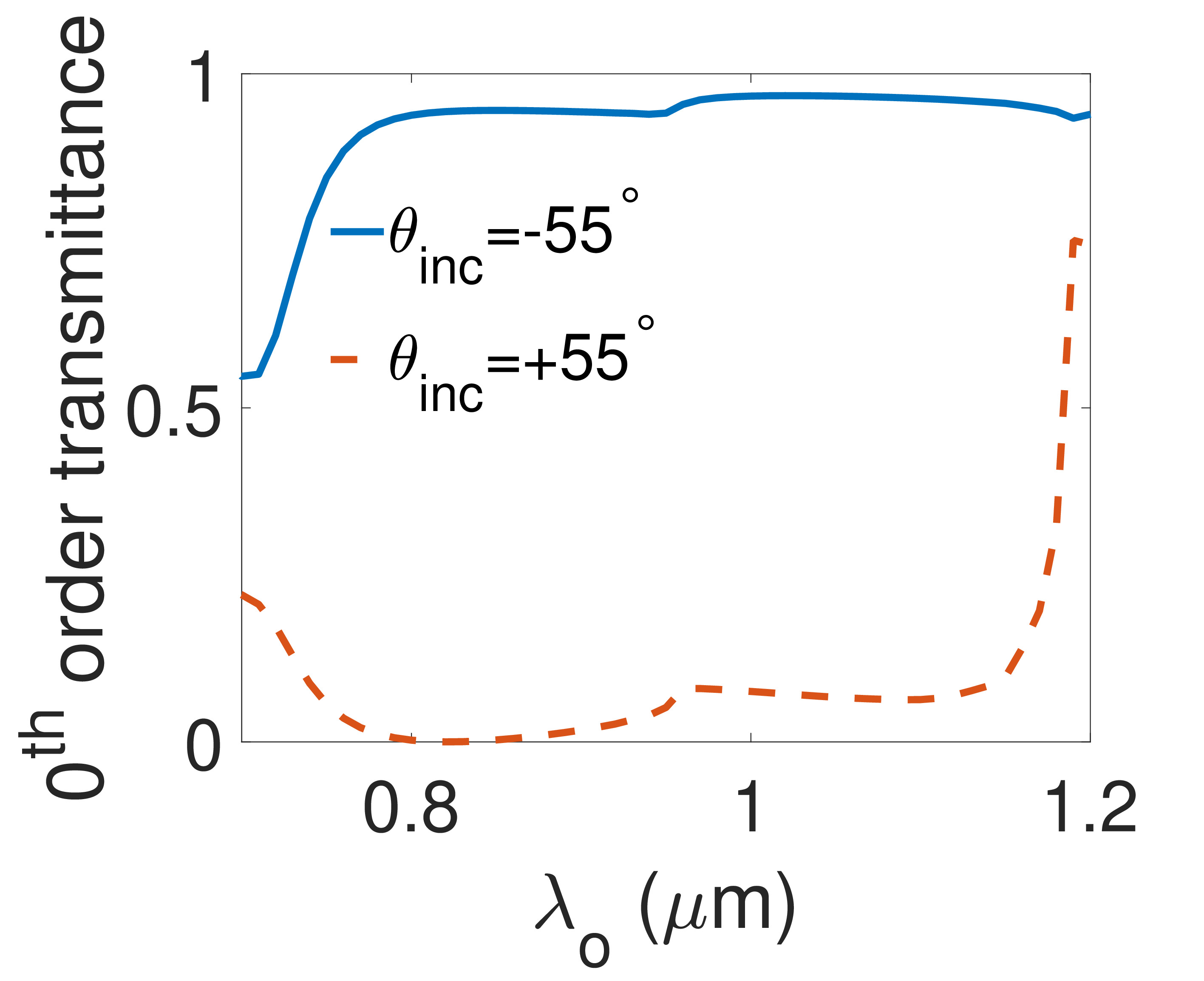}
         \caption{}
    
     \end{subfigure}
\begin{subfigure}[b]{0.3\textwidth}
         \includegraphics[width=\textwidth]{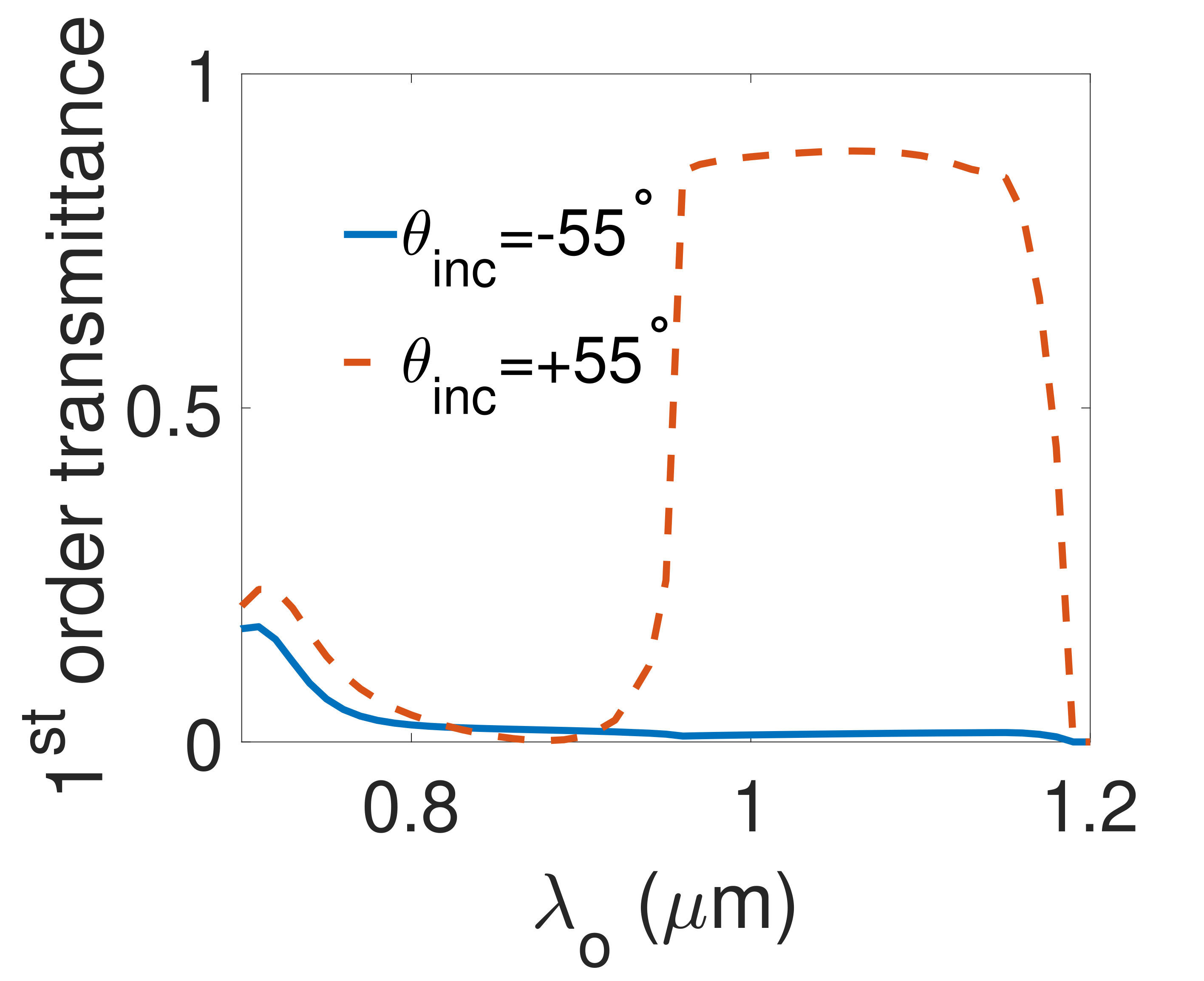}
         \caption{}
  
     \end{subfigure}
\begin{subfigure}[b]{0.3\textwidth}
         \includegraphics[width=\textwidth]{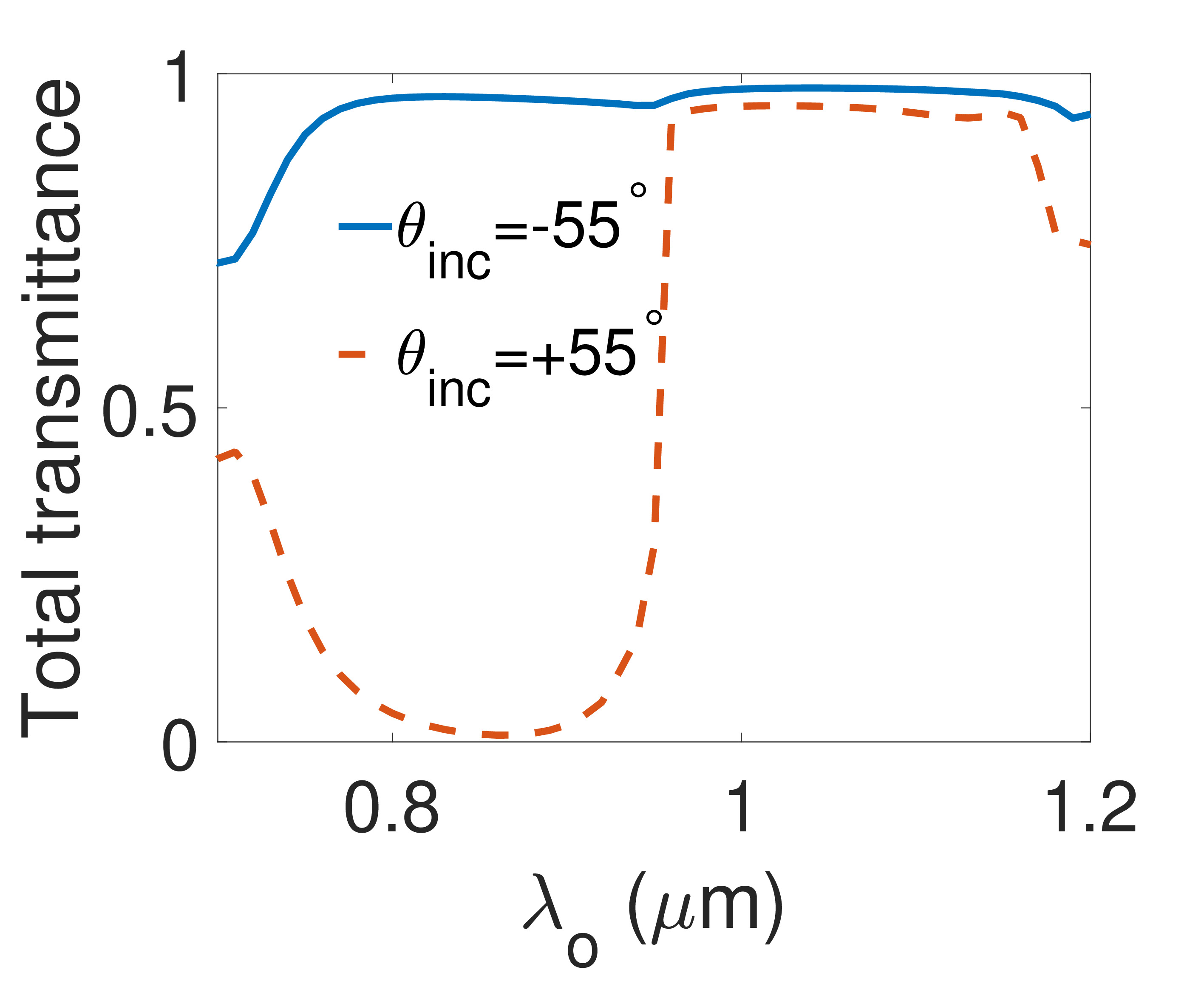}
         \caption{}

     \end{subfigure} 
\begin{subfigure}[b]{0.3\textwidth}
         \centering
         \includegraphics[width=\textwidth]{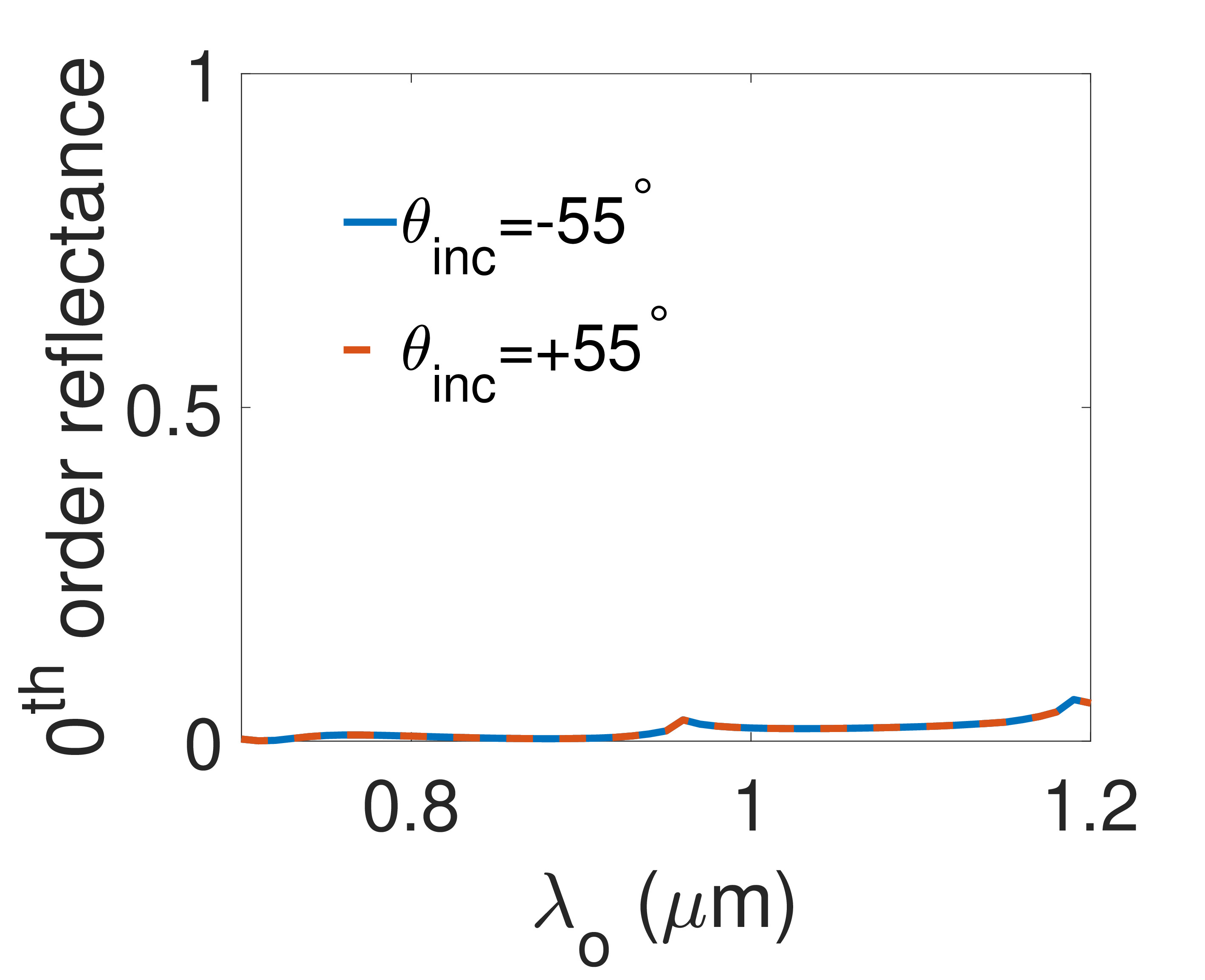}
         \caption{}
     
     \end{subfigure}
\begin{subfigure}[b]{0.3\textwidth}
         \includegraphics[width=\textwidth]{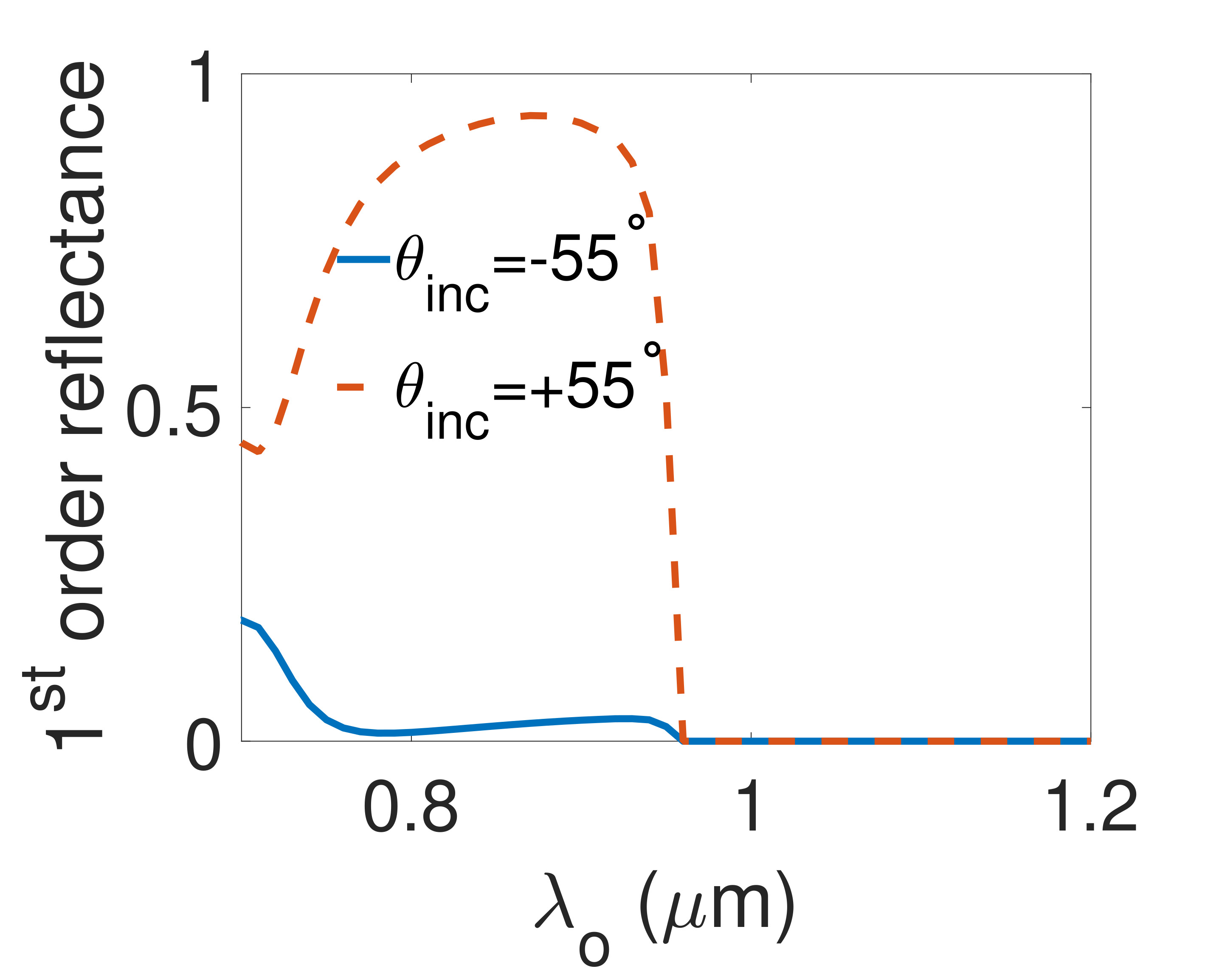}
         \caption{}
        
     \end{subfigure}
\begin{subfigure}[b]{0.3\textwidth}
         \includegraphics[width=\textwidth]{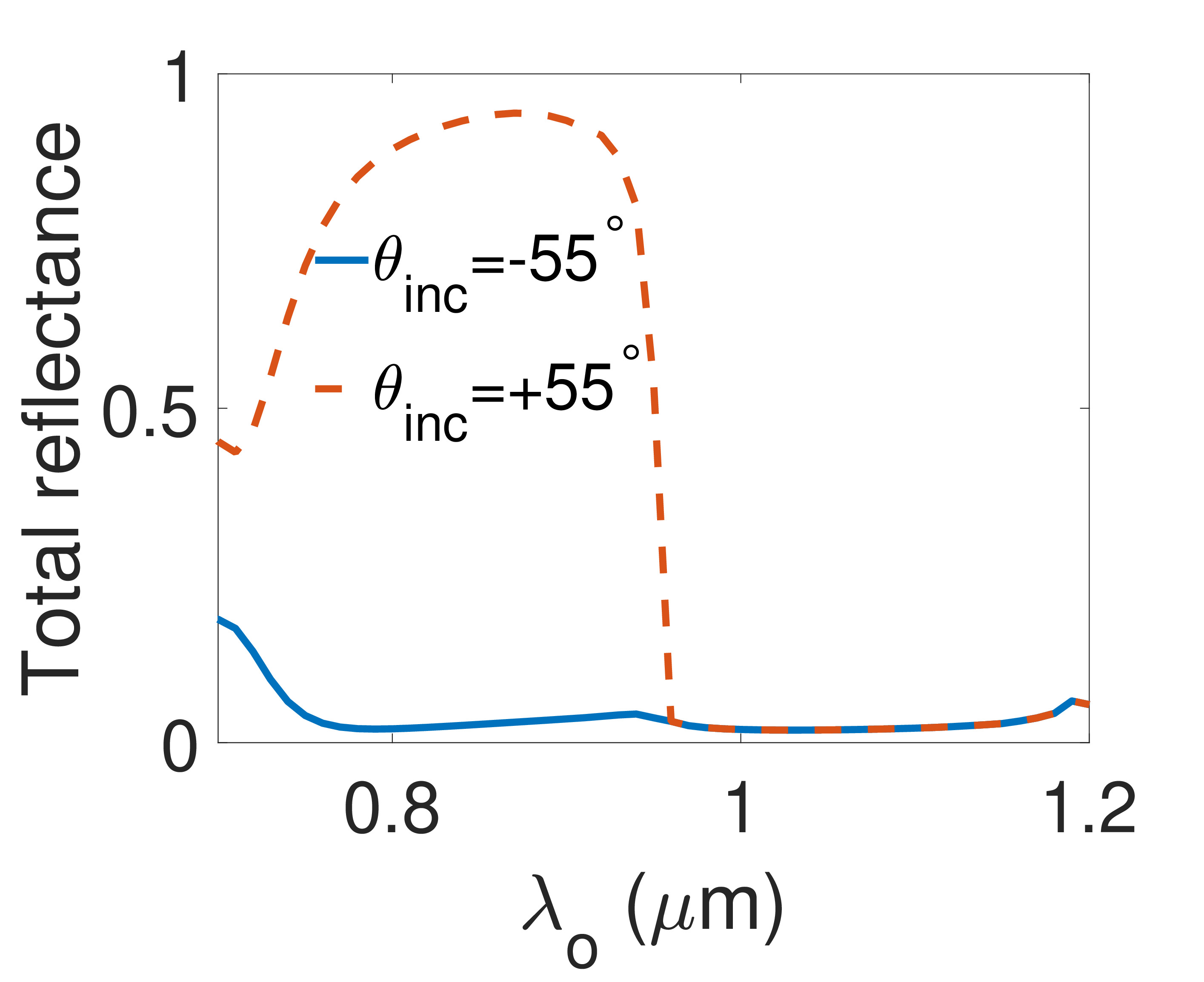}
         \caption{}
       
     \end{subfigure}
     \caption{(a) $0^{\rm th}$-order, (b) 1$^{\rm st}$-order, and (c) overall transmittance of the introduced metagrating for two angles of incidence $-55^{\circ}$ and $55^{\circ}$. (d) 0$^{\rm th}$-order, (e) 1$^{\rm st}$-order, and (f) overall reflectance.}
 \label{fig:TR}    
\end{figure}

The $0^\text{th}$-order, $1^\text{st}$-order, and overall transmittance of the introduced metagrating are shown in Fig.~\ref{fig:TR} (a), (b), and (c), respectively, for a pair of incidence angles equal to $-55^{\circ}$ and $55^{\circ}$. A notable asymmetry  between the two incidence angles in the $0^\text{th}$-order and overall transmittance is present for operation bandwidths beginning from 0.8 to 1.18 $\rm \mu m$, and from 0.8 to 0.95 $\rm \mu m$, respectively. The physical origin of such symmetry breaking is rooted in the scattering of the oblique gold strips, which is engineered to 
reflect the input energy into the $1^\text{st}$ diffraction order for positive angles of incidence. The $0^\text{th}$-order, $1^\text{st}$-order, and overall reflectance of the metagrating are also shown in Fig.~\ref{fig:TR} (d), (e), and (f), respectively. Reciprocity theorem enforces the $0^\text{th}$-order reflectance to be symmetric as shown in Fig.~\ref{fig:TR} (d). Nevertheless, the scattering response of the gold strips is optimized to minimize the specular reflectance. Consequently, the angular transmittance is remarkably asymmetric in the diffractive regime without imposing angularly asymmetric absorbance on the system.

\begin{figure}[!htb]
\centering
\begin{subfigure}[b]{0.12\textwidth}
         \centering
         \includegraphics[width=\textwidth]{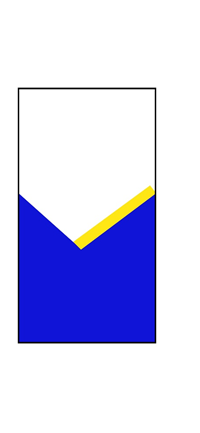}
         \caption{}
   
     \end{subfigure}
\begin{subfigure}[b]{0.27\textwidth}
         \centering
         \includegraphics[width=\textwidth]{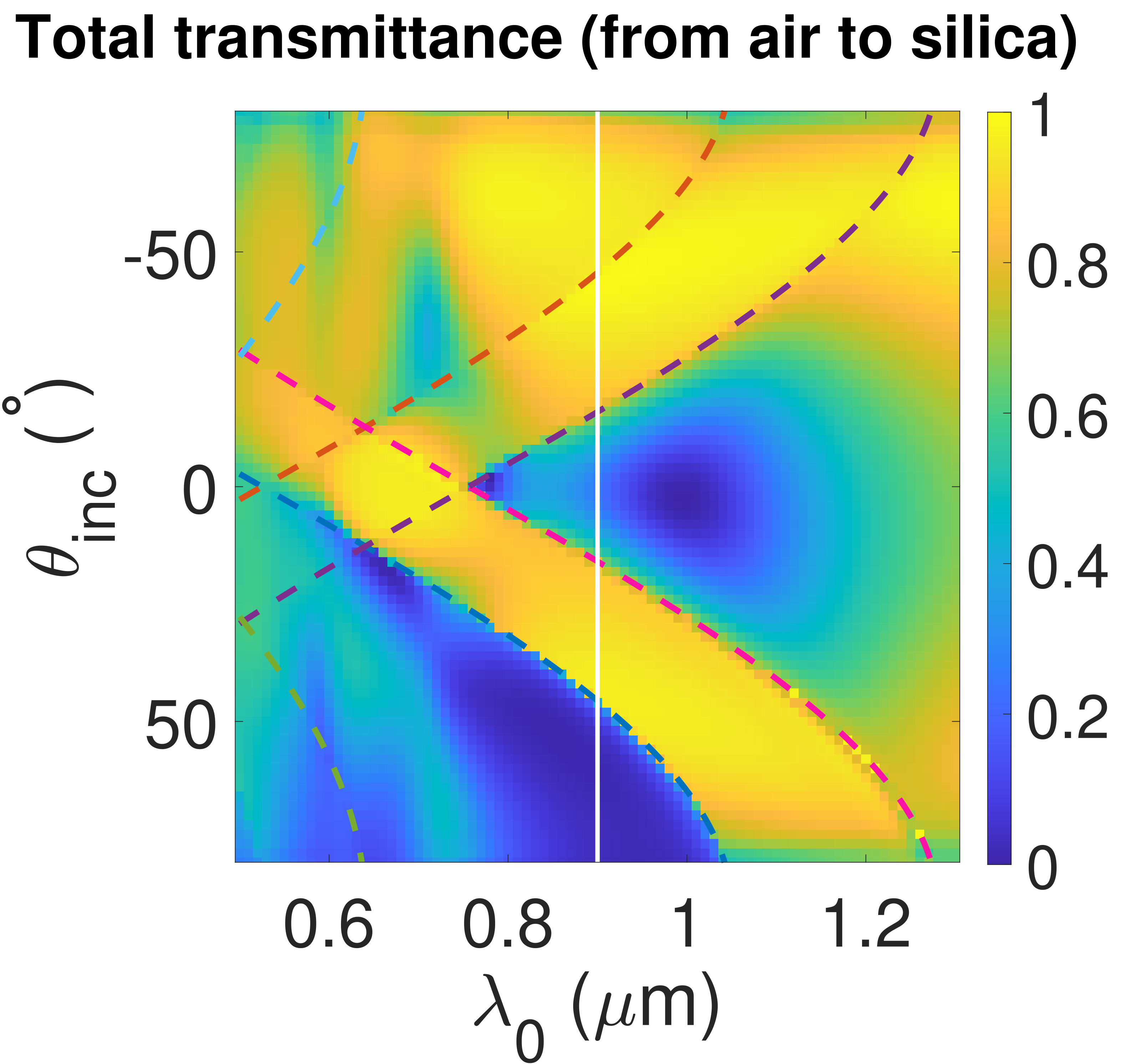}
         \caption{}
     
     \end{subfigure}
\begin{subfigure}[b]{0.27\textwidth}
         \includegraphics[width=\textwidth]{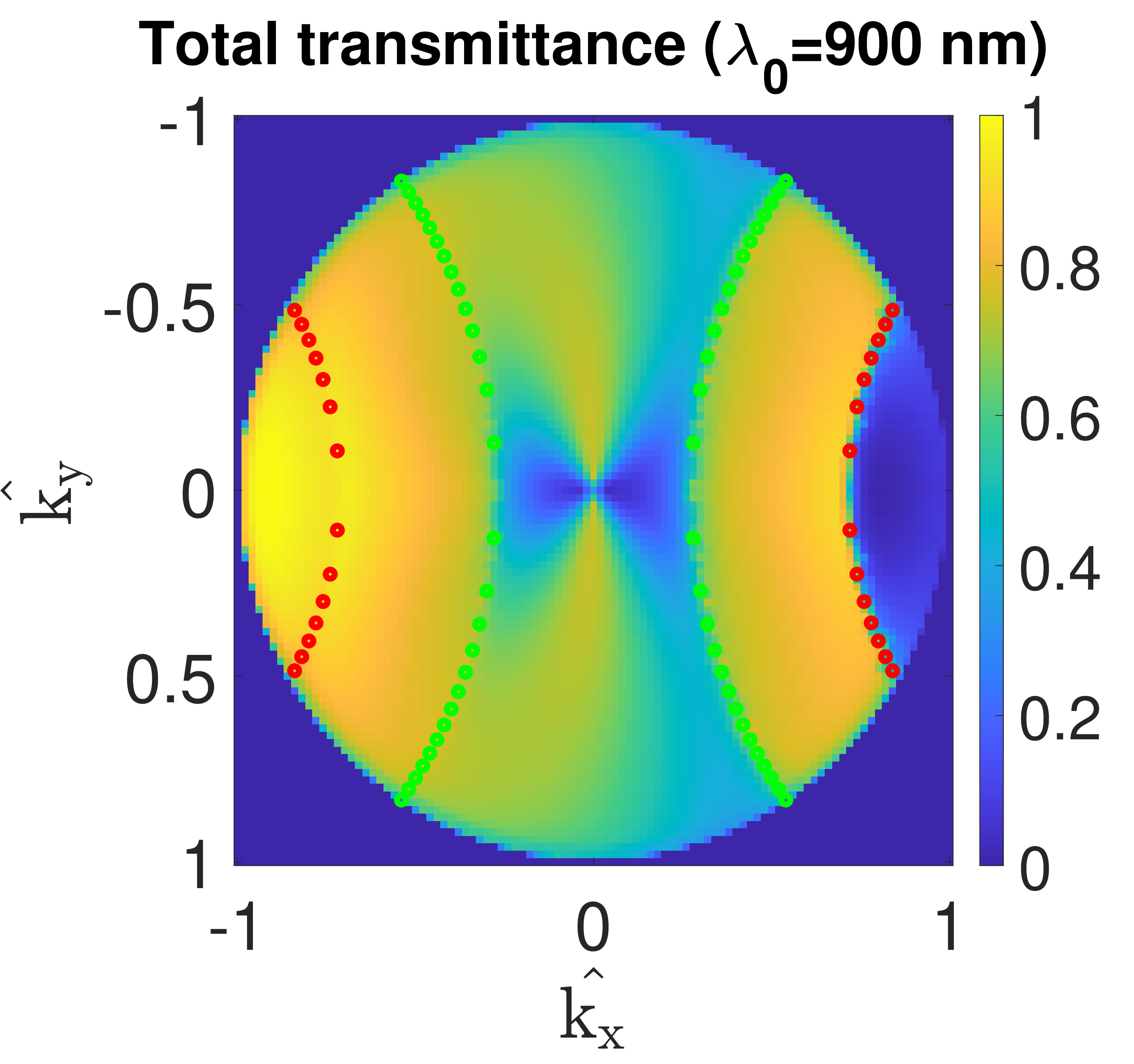}
         \caption{}
       
     \end{subfigure}
\begin{subfigure}[b]{0.27\textwidth}
         \includegraphics[width=\textwidth]{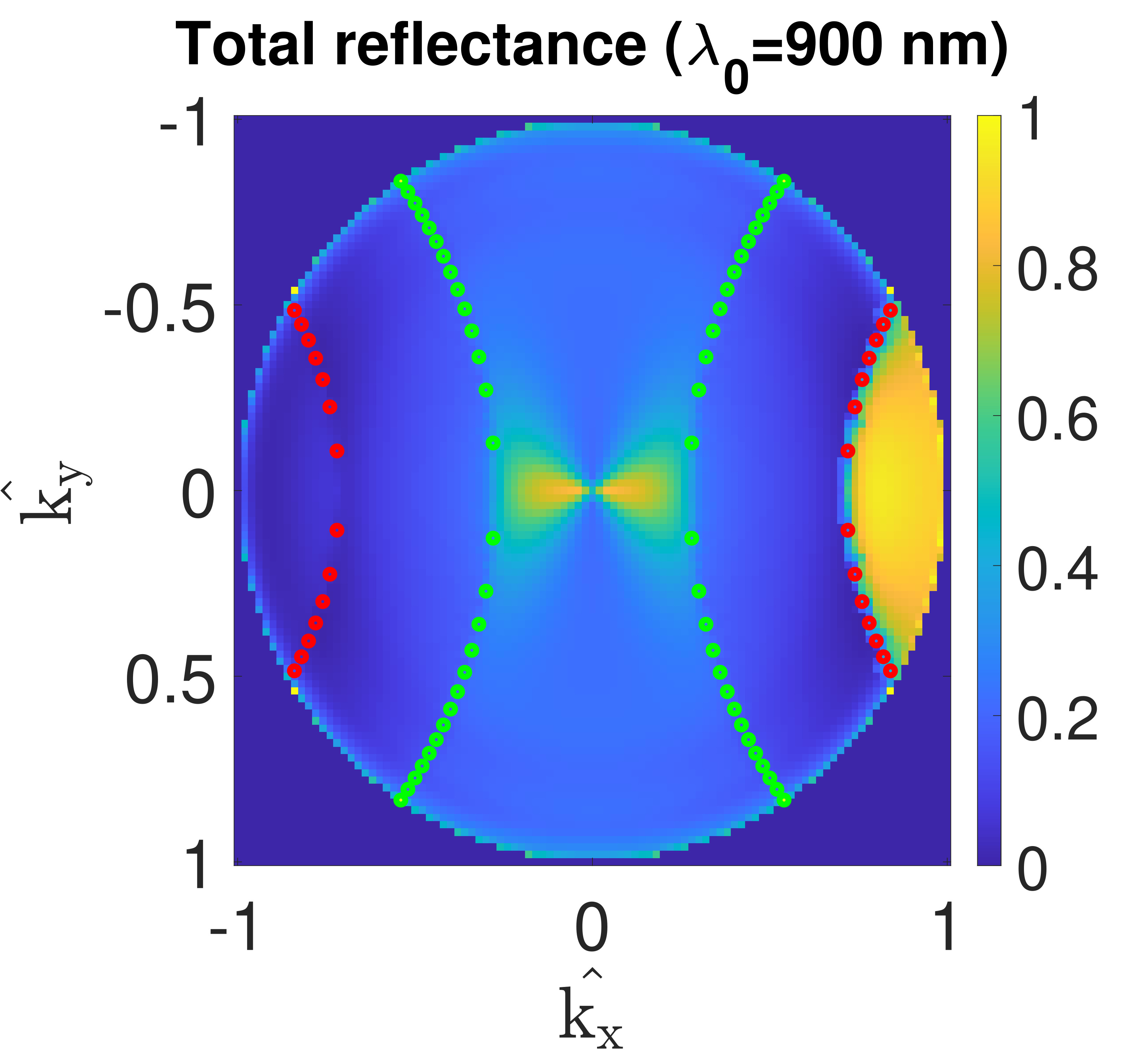}
         \caption{}
     
     \end{subfigure} 

\begin{subfigure}[b]{0.12\textwidth}
         \centering
         \includegraphics[width=\textwidth]{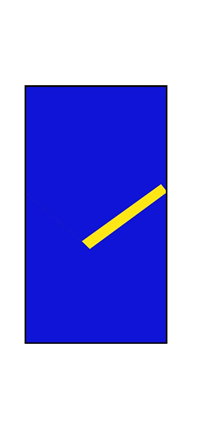}
         \caption{}
      
     \end{subfigure}   
\begin{subfigure}[b]{0.27\textwidth}
         \centering
         \includegraphics[width=\textwidth]{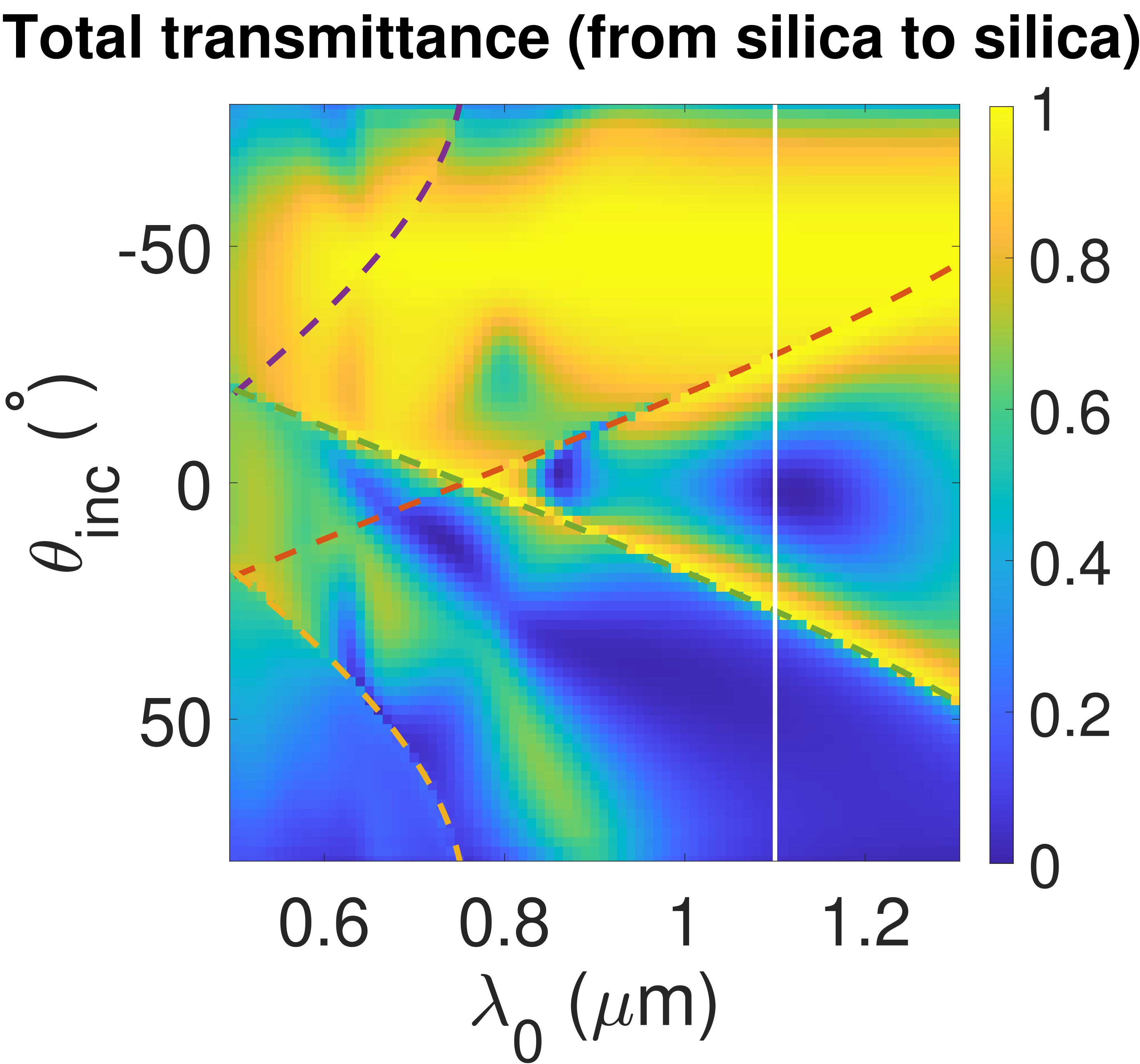}
         \caption{}
      
     \end{subfigure}
\begin{subfigure}[b]{0.27\textwidth}
         \includegraphics[width=\textwidth]{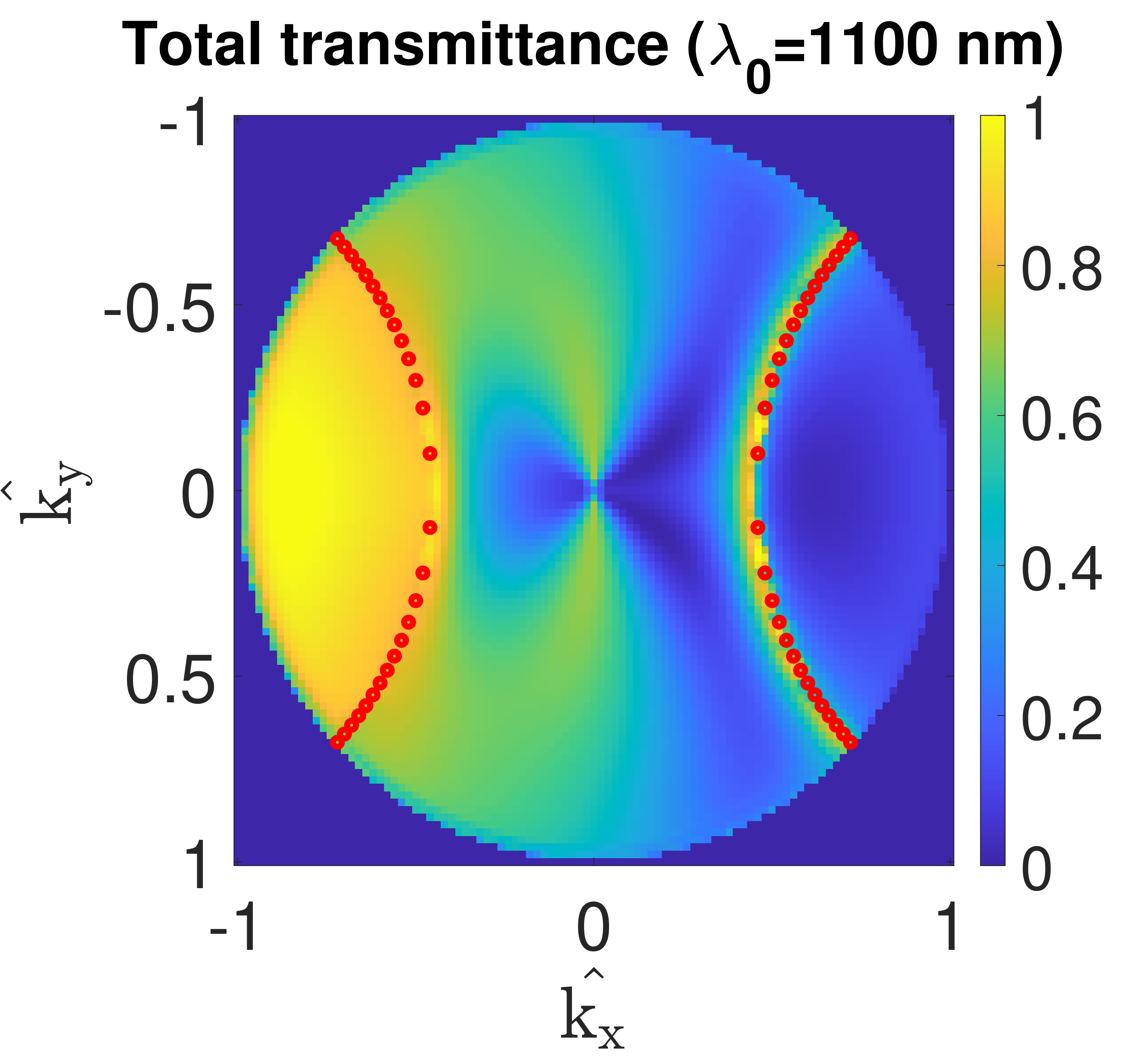}
         \caption{}
      
     \end{subfigure}
\begin{subfigure}[b]{0.27\textwidth}
         \includegraphics[width=\textwidth]{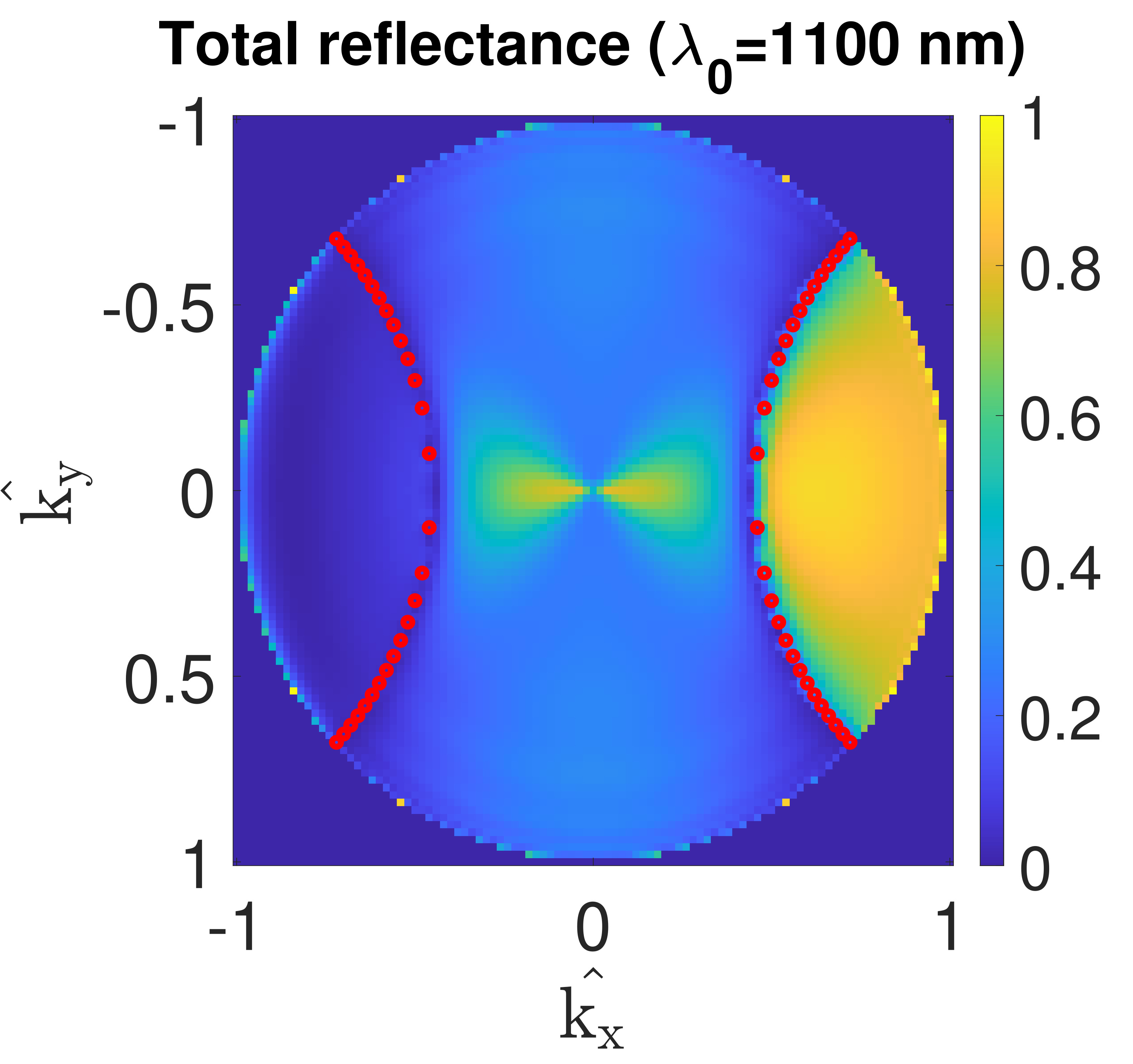}
         \caption{}
       
     \end{subfigure}
     \caption{(a) Schematic representing a case for incidence from air to silica. (b) The overall transmittance as a function of $\lambda_{0}$ and $\theta$ for the incident wave. (c) Overall transmittance and (d) reflectance for all of the spatial Fourier space for $ \lambda_{\rm 0}=900$ nm. (e) A schematic depicting another case for incidence when all of the surrounding medium is made of silica. (f), (g), and (f) analogous to (b), (c), and (d), respectively, for the schematic diagram shown in (e) ($ \lambda_{\rm 0}=1100$ nm). The dashed lines correspond to the propagation conditions of diffraction orders in the reflection and transmission domains.}
     \label{fig:momentum}
\end{figure}

To visualize the angular scattering behavior of the metagrating, we now study its response in Fourier space for spatial frequencies along both $x$ and $y$ axes. This means that the plane of incidence for the input TM plane waves is not restricted to the plane of translation symmetry (i.e., the $xz$-plane). Consequently, the directions of diffracted waves are no longer restricted to this plane either and, owing to momentum conversation, the locus of the diffraction orders becomes a conical surface. This phenomenon, known as conical diffraction, is typically studied by finding the exact solutions of the electromagnetic fields of an arbitrary grating based on Maxwell's equations~\cite{conical}. Here, we consider a simpler alternative based on symmetry considerations (see a detailed discussion in Supplementary Section 2) to find the cutoff condition in Fourier space for each diffraction order to start propagating. The resulting condition is given by 
\begin{equation}
     \hat{k}_y^2=\dfrac{n_{\rm o}^2}{n_{\rm i}^2}-\left(\hat{k}_x+\dfrac{m\lambda_{\rm 0}}{n_{\rm i}\Lambda}\right)^2,
    \label{eq:cutoff_moment}
\end{equation}
where $\hat{k}_x$ and $\hat{k}_y$ are the normalized components of the wave vector in the $x$ and $y$ directions, whereas $\Lambda$, $\lambda_{0}$, $n_\text{o}$, and $n_\text{i}$ denote the grating period, the free-space wavelength, the input and output port refractive indices, respectively.
Figure~\ref{fig:momentum} (a) depicts a unit cell of the proposed metagrating when the input wave is incident from air to silica. Figure~\ref{fig:momentum} (b) shows the total transmittance as a function of the incidence angle and the input wavelength for the case shown in  Fig.~\ref{fig:momentum} (a). Note that there are four curves around which abrupt changes in transmittance occur. These curves denote the cutoff conditions for the first diffraction order in the reflection and transmission media, which are derived from the grating equation. 
Additionally, it can be seen that there is remarkable angular asymmetry in the total transmittance between negative and positive angles greater than roughly $40^{\circ}$ spanning an operation bandwidth starting from 0.8 to 0.95 $\rm \mu m$).
For $\lambda_{0}=0.9\ \mu m$, the total transmittance is computed for all of the Fourier space as shown in Fig.~\ref{fig:momentum} (c). This is achieved by sampling a sphere representing all of the possible wave vector components for the incident plane wave, while computing the transmittance for every sample based on full-wave simulations. 
It can also be noted that there are four lines representing the cutoff conditions for the first diffraction order in transmission and reflection domains as described by Eq.~\ref{eq:cutoff_moment}. The total reflectance is also shown in Fig.~\ref{fig:momentum} (d) which implies that the metagrating has theoretically low absorption at this wavelength when summed to the total transmittance in Fig.~\ref{fig:momentum} (c). 

For comparison, we also consider another unit cell that is depicted in Fig.~\ref{fig:momentum} (e) for the case when the reflection and transmission domains are both made of silica (gold strips embedded into silica).
Figures~\ref{fig:momentum} (f), (g), and (h) are analogous to Figs.~\ref{fig:momentum} (b), (c), and (d), respectively, depicting the angular dispersion of the total transmittance, total transmittance and reflectance as a function of the spatial frequencies in the $x$ and $y$ directions at a wavelength equal to 1.1$\mu m$ for the unit cell shown in Fig.~\ref{fig:momentum} (e). Note that the physical difference between this unit cell and the one in Fig.~\ref{fig:momentum} (a) is the fact that the propagation conditions for the first diffraction order in the reflection and transmission regimes (i.e., the four curves in Fig.~\ref{fig:momentum} (b)) merge in two curves leading to a broader range for angular asymmetry.

For both structures in Figs.~\ref{fig:momentum} (a) and (b), there is a minor amount of angular asymmetry in the non-diffractive regime. This can be explained by the fact that gold exhibits ohmic losses, which results in angularly asymmetric absorption ($A(\theta) \neq A(-\theta)$) due to the broken symmetry of the array.

\begin{figure}
\centering
\begin{subfigure}[b]{0.485\textwidth}
\includegraphics[width=\textwidth]{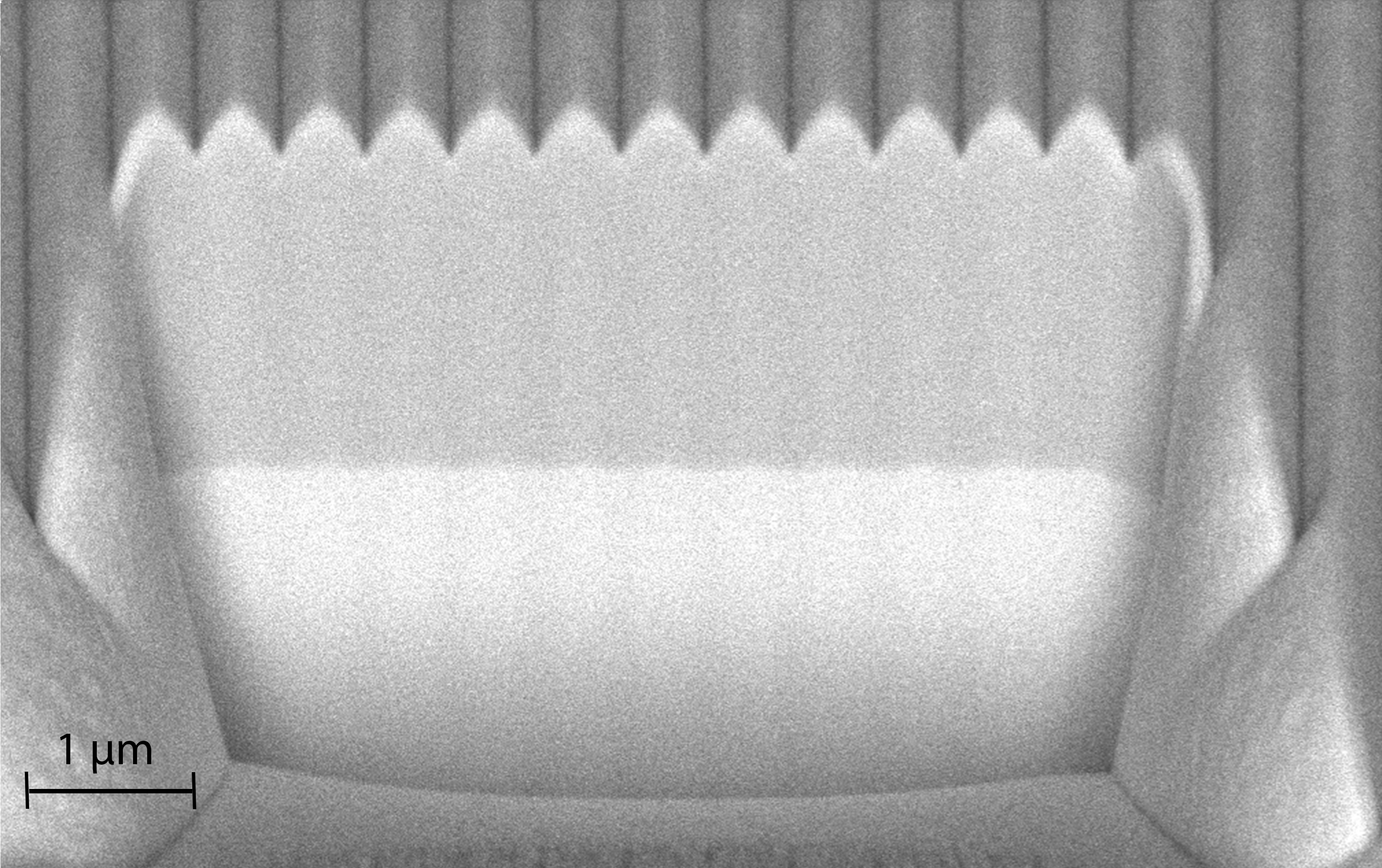}
\caption{}
\label{fig:loss_forward}
\end{subfigure}
\begin{subfigure}[b]{0.452\textwidth}
\includegraphics[width=\textwidth]{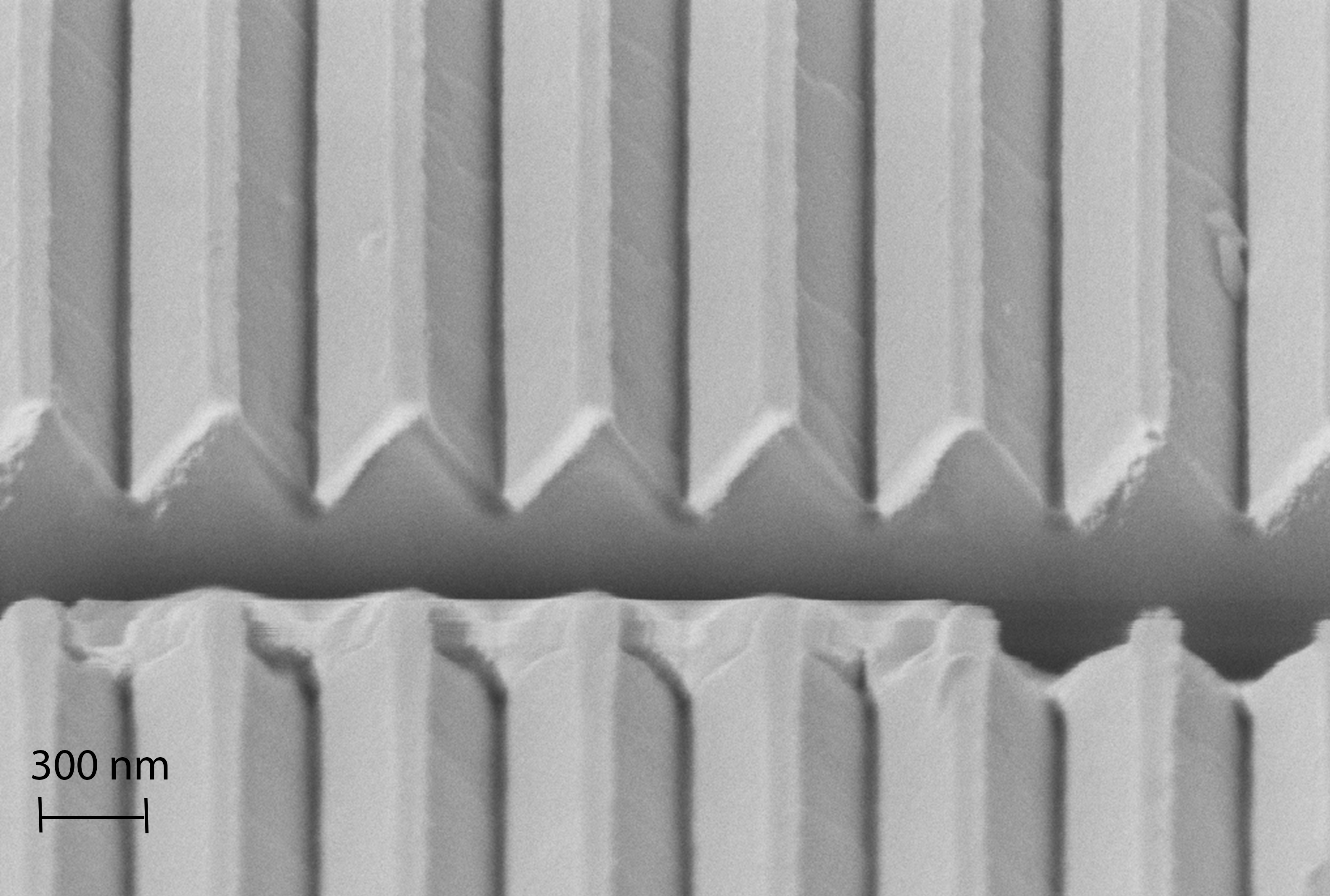}
\caption{}
\end{subfigure}
\caption{(a) SEM image of the cross section of the fabricated triangular silica ridges after thermal oxidation. (b) Same after oblique evaporation of gold on a trial sample.}
\label{fig:fab_sample}
\end{figure}

We have fabricated and optically characterized a sample of the proposed metagrating structure with an area equal to 1 $\rm cm^{2}$. The process flow for fabrication is described in the Supplementary Section 6. The scanning electron micsocope (SEM) image in figure~\ref{fig:fab_sample} (a) shows a cross-sectional view of the fabricated sample (obtained using focused-ion-beam (FIB) milling), after thermal wet oxidation processing of silicon. Note that the interface between silicon and silica below the periodic array is smooth and non-diffractive. Moreover, the gold layer is evaporated on only one side of a triangular pattern as demonstrated in Fig.~\ref{fig:fab_sample} (b) for a trial sample before oxidation (an SEM image of the investigated sample is shown in Fig. S9 the supplemental materials).

\begin{figure}[!htb]
\centering
\begin{subfigure}[b]{0.45\textwidth}
         \centering
         \includegraphics[width=\textwidth]{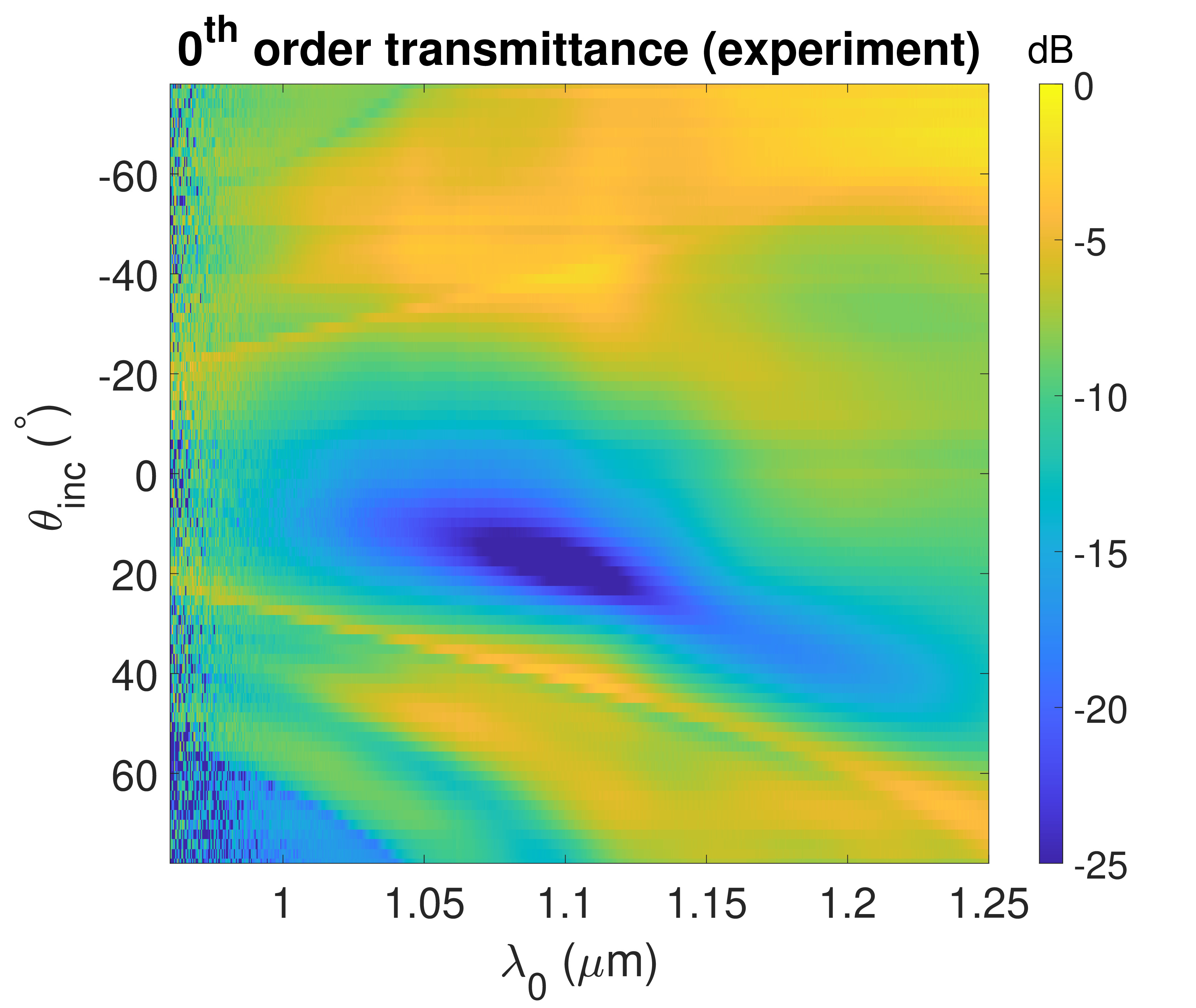}
         \caption{}
         
     \end{subfigure}
\begin{subfigure}[b]{0.45\textwidth}
         \includegraphics[width=\textwidth]{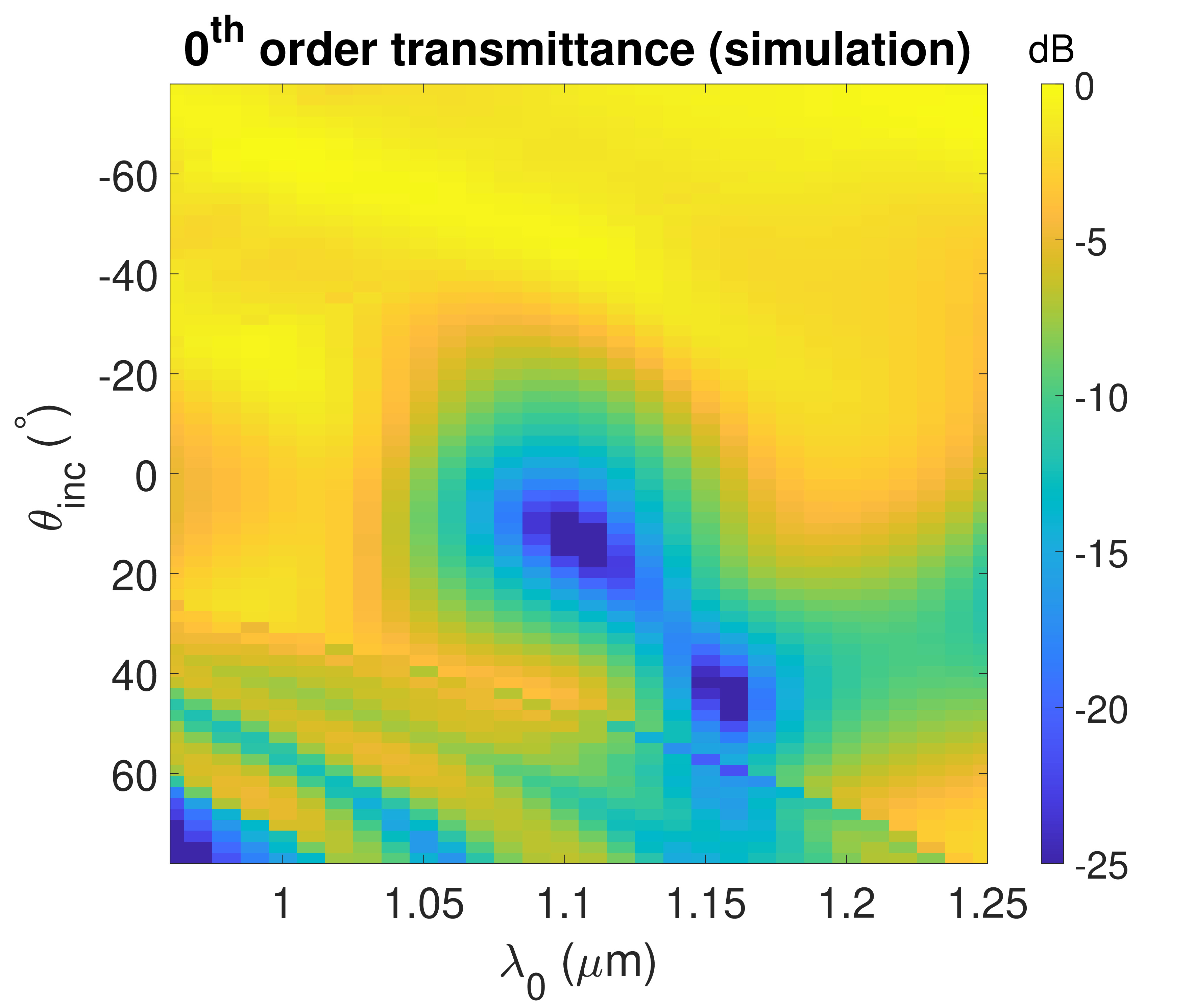}
         \caption{}
         
     \end{subfigure} 
\label{fig:meta_pinhole}
\centering

     \caption{(a) $0^{\rm th}$-order transmittance measured by the experimental setup as a function of the incidence angle of the TM-polarized light source.(b) Similar as (a) but based on full-wave simulations.}
     \label{Fig:exp_measurements}
\end{figure}

The measured angular dispersion of the $0^\text{th}$-order transmittance, for the fabricated sample of the metagrating, is shown in Fig.~\ref{Fig:exp_measurements} (a). The experimental measurements of the $0^\text{th}$-order transmittance are normalized with respect to that of a blank silicon wafer polished on both sides. It should be noted that the symmetry of the $0^\text{th}$-order angular transmittance is broken both in the non-diffractive and diffractive region, which leads to a broader angular asymmetry range. In the non-diffractive regime, the asymmetry in transmittance exceeds 20 dB around a central wavelength of 1.1 $\mu m$. Fig.~\ref{Fig:exp_measurements} (b) shows the simulated $0^\text{th}$-order transmittance calculated by a finite-element solver. Note that the simulation results follow a similar trend to that of the experimental measurements in Fig.~\ref{Fig:exp_measurements} (a). However, there are some differences, particularly, in the non-diffractive regime due to fabrication imperfections and the material loss, which is not considered in the simulations. The most noticeable is the strong dependence of the angular absorption on the curvature of the gold strips. As it can be seen in Fig. S7 (b) and (d) in the Supplemental Materials, the location of the transmission dip shifts remarkably for different curvature radii in the $x-$ and $z-$directions.
The optical measurements where performed using  a super-continuum white light source (SuperK Extreme EXW-12 by NKT photonics), illuminating the sample, which is placed on a rotating holder. The beam is nearly collimated and is approximately 3 mm in diameter, which mimicks plane-wave like excitation. The $0^\text{th}$-order transmittance was recorded as a function of angle of incidence using an optical spectrum analyzer (YOKOGAWA AQ6370D). Since the positions of the source and the detector are fixed, only the zero diffraction order is measured in transmission,  as it is always collinear with the incident beam axis after going through the sample. Refer to the Supplementary Section 8 for the additional information regarding the optical setup. 

\section{Conclusion}
In this work, we describe how an engineered diffractive optical system in the form of a metagrating enables a virtual knife-edge in Spatial Fourier Space. The metagrating can switch from transmission to reflection in an abrupt fashion, depending on the incidence angle. Such a response can find several applications in imaging and spatial computing applications such as achieving image filtering or a step function in Fourier space in an ultra-thin compact fashion.
 

\pagebreak
\input{osa-supplemental-document-template}

\end{document}

%% file: osa-supplemental-document-template.tex


\title{Supplemental Material: Step Function in momentum space by a metagrating}
\author{} 




\section{Symmetry analysis}

In what follows, we investigate the scattering properties of a diffraction grating in terms of spatial symmetries. For simplicity, we restrict our analysis to a diffraction grating that periodically varies along the $x$ axis and whose period is such that only the zeroth and first diffraction orders exist for a given set of incidence angle, free space wavelength and background refractive index (assumed to be identical on both sides of the grating). We also consider the grating to be infinite along the $y$ direction.

\begin{figure}[h!]
\centering
\includegraphics[width=0.8\textwidth]{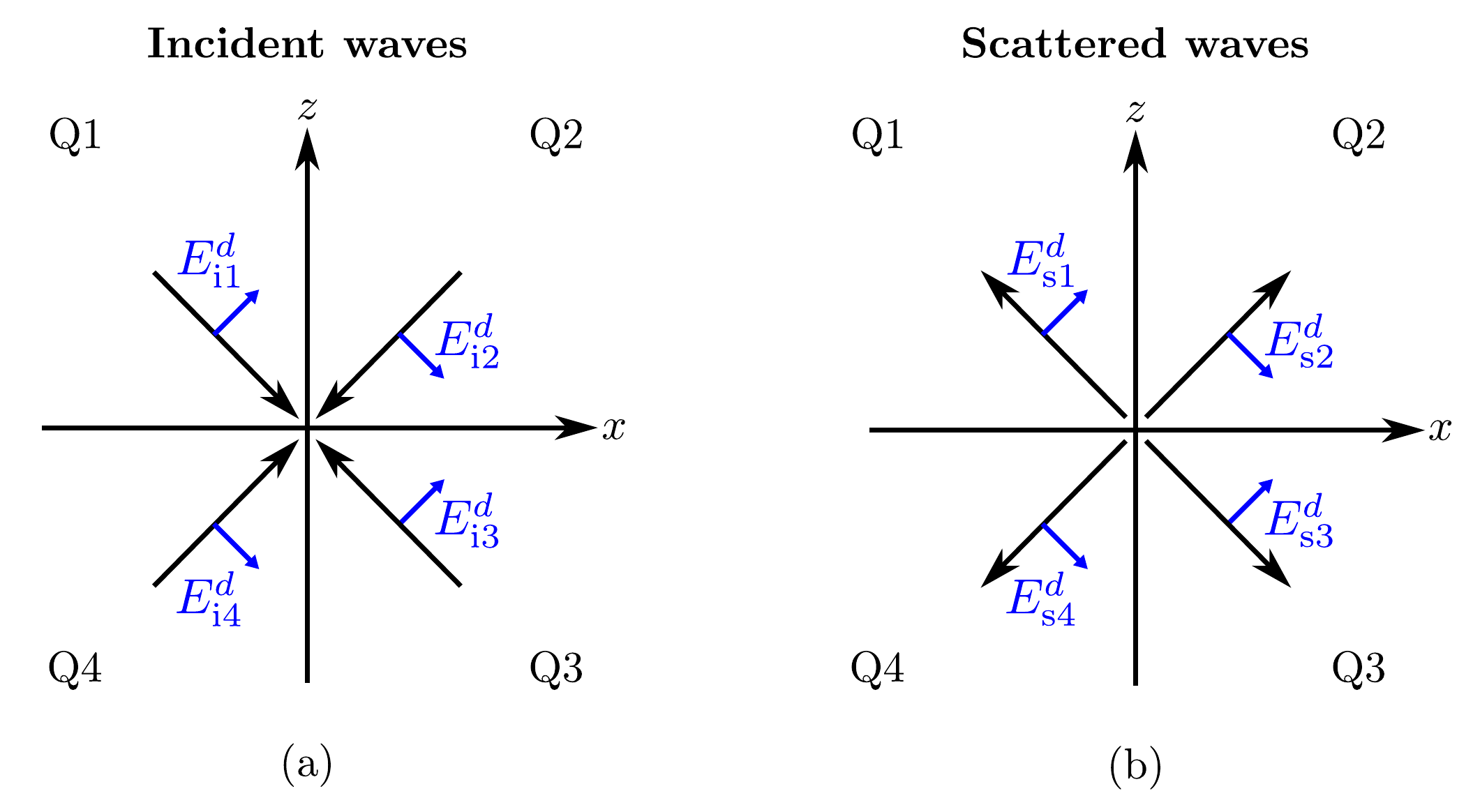}
\caption{Naming and polarization convention for the fields in the incident and scattering channels of all four quadrants denoted Q1 to Q4.}
\label{fig:schematic_DO}
\end{figure}

From a general perspective, we can express the scattering matrix, $\te{S}$, of this grating as
\begin{equation}
	\label{eq_GSM}
	\ve{E}_\text{s} = \te{S}\cdot\ve{E}_\text{i},
\end{equation}
where $\ve{E}_\text{i}$ and $\ve{E}_\text{s}$ are vectors containing the complex amplitude of the incident and scattered waves, respectively. For our simplified system,~\eqref{eq_GSM} reduces to
\begin{equation}
 \label{Eq:symscat}
\begin{bmatrix}
\ve{E}_\text{s}^0\\
\ve{E}_\text{s}^1
\end{bmatrix}=
\begin{bmatrix}
\te{S}^{00} & \te{S}^{01} \\
\te{S}^{10} & \te{S}^{11} 
\end{bmatrix}\cdot
\begin{bmatrix}
\ve{E}_\text{i}^0\\
\ve{E}_\text{i}^1
\end{bmatrix},
\end{equation}
where the complex amplitude vectors for the incident and scattered orders are given by
\begin{equation}
	\ve{E}_\text{a}^d=[E_\text{a1}^d\,\, E_\text{a2}^d\,\, E_\text{a3}^d\,\, E_\text{a4}^d]^T,
\end{equation}
where $T$ denotes the transpose operation. The amplitudes $E_\text{an}^d$, with $n=\{1,2,3,4\}$ being the quadrant number, $d=\{0,1\}$ the channel number and $a=\{\text{i},\text{s}\}$, correspond to waves incident or scattered on/by the grating from all four quadrants of the $xz$ plane, as depicted in Fig.~\ref{fig:schematic_DO}.
Note that while we only consider TM polarized waves in what follows, our analysis is identically valid for TE polarized waves. Additionally, the fact that the considered grating is invariant along the $y$ direction implies that polarization conversion between TM and TE waves is impossible. 

In this formalism, the channels correspond to directions (propagation angles) in which the energy is allowed to flow based on the grating equation in~\eqref{eq_gratingEq}. To fully construct the scattering matrix, $\te{S}$, we consider that both the incident and scattered waves may use the same channels. This means, for instance, that the incident wave may impinge on the grating from an angle that would otherwise correspond to a 1\textsuperscript{st}-order diffracted wave. It follows that the sub-scattering matrices, $\te{S}^{mn}$, in~\eqref{Eq:symscat} may take either of the following forms
\begin{equation}
\label{eq_Smn}
\te{S}^{mn}_\text{F} = 
 \begin{bmatrix}
0 & S_{12}^{mn} & S_{13}^{mn} & 0 \\
S_{21}^{mn} & 0 & 0 & S_{24}^{mn} \\
S_{31}^{mn} & 0 & 0 & S_{34}^{mn} \\
0 & S_{42}^{mn} & S_{43}^{mn} & 0 
\end{bmatrix}
\qquad
\text{or}
\qquad
\te{S}^{mn}_\text{B}=
\begin{bmatrix}
S_{11}^{mn} & 0 & 0 & S_{14}^{mn} \\
0 & S_{22}^{mn} & S_{23}^{mn} & 0 \\
0 & S_{32}^{mn} & S_{33}^{mn} & 0 \\
S_{41}^{mn} & 0 & 0 & S_{44}^{mn} \\
\end{bmatrix},
\end{equation}
where the subscript `$\text{F}$' indicates \emph{forward} scattering (i.e., when $\text{sign}(k_{x,\text{i}})=\text{sign}(k_{x,\text{s}})$, whereas the subscript `$\text{B}$' indicates \emph{backward} scattering (i.e., when $\text{sign}(k_{x,\text{i}})=-\text{sign}(k_{x,\text{s}})$ with $k_{x,\text{i}}$ and $k_{x,\text{s}}$ being the $x$-components of the k-vector of the incident and scattered waves, respectively. The components of the matrices in~\eqref{eq_Smn} correspond to the scattering parameters between different quadrants. For instance, $S_{12}^{01}$ corresponds to the scattering coefficient between an incident wave propagating along the +1 diffraction channel (corresponding to the angle of the first diffraction order) of the second quadrant (Q2) and a scattered wave propagating along the 0\textsuperscript{th} diffraction channel (corresponding to the angle of the 0\textsuperscript{th} diffraction order) of the first quadrant (Q1). For the specific grating considered in Fig.~2 of the main text, the grating parameters are such that $\te{S}^{00}$ and $\te{S}^{11}$ take the shape of forward scattering matrices, whereas $\te{S}^{01}$ and $\te{S}^{10}$ are backward scattering matrices. This is because the grating is designed for retro-reflection implying that the reflection between a 0\textsuperscript{th}-order channel and 1\textsuperscript{st}-order channel occurs within the same quadrant, as depicted in the right-hand side of Fig.~2 of the main text.

Following a similar procedure as in~\cite{dmitriev2013a,achouri2023a}, we now express the effect of a spatial symmetry operation $\Lambda$, given by the matrix $\te{M}_\Lambda$, onto the fields via
\begin{equation}
\ve{E}_\text{a}^{d'} = \te{M}_\Lambda\cdot\ve{E}_\text{a}^d,
\end{equation}
where $\ve{E}_\text{a}^{d'}$ corresponds the transformed $\ve{E}_\text{a}^{d}$. Since, in our case, the geometry of the grating is restricted to a two-dimensional problem in the $xz$ plane, it follows that the only spatial symmetry operations that are possible are $\sigma_x$ (reflection through the $x$ axis), $\sigma_z$ (reflection through the $z$ axis) and $C_{2y}$ (180$^\circ$-rotation symmetry along the $y$ axis). The corresponding transformation matrices, that may be obtained by intuitively inspecting how these symmetry operations affect the fields in Fig.~\ref{fig:schematic_DO}, are given by
\begin{equation}
    \te{M}_{\sigma_x} =
\begin{bmatrix}
0 & 1 & 0 & 0 \\
1 & 0 & 0 & 0 \\
0 & 0 & 0 & 1 \\
0 & 0 & 1 & 0 
\end{bmatrix},\quad
\te{M}_{\sigma_z} =
\begin{bmatrix}
0 & 0 & 0 & 1 \\
0 & 0 & 1 & 0 \\
0 & 1 & 0 & 0 \\
1 & 0 & 0 & 0 
\end{bmatrix},\quad
\te{M}_{C_{2y}}=
\begin{bmatrix}
0 & 0 & 1 & 0 \\
0 & 0 & 0 & 1 \\
1 & 0 & 0 & 0 \\
0 & 1 & 0 & 0 
\end{bmatrix}.
\end{equation}
It directly follows that if the diffraction grating is \emph{invariant} under one of these symmetry operations, then its sub-scattering matrices must satisfy the symmetry invariance condition
\begin{equation}
\label{eq_syminv}
    \te{S}^{mn}\cdot\te{M}_\Lambda = \te{M}_\Lambda\cdot\te{S}^{mn}.
\end{equation}
Additionally, we are considering that the grating is reciprocal implying that $\te{S} = \te{S}^T$ and, consequently, that
\begin{equation}
\label{eq_recip}
	\te{S}^{mm} = \te{S}^{mm^T} \quad \text{and} \quad \te{S}^{mn} = \te{S}^{nm^T}.
\end{equation}

Applying this formalism to the two diffraction gratings in Fig.~2 of the main text, implies solving~\eqref{eq_syminv} along with~\eqref{eq_recip} considering that the grating in the top row is symmetric under $\sigma_z$, which leads to
\begin{equation}
    \bar{\bar{S}}^{\rm 00}=\begin{bmatrix} 0 & \mathcolor{magenta}{S_{12}^{00}} &\mathcolor{blue}{S_{13}^{00}} &\rm 0\\\mathcolor{magenta}{S_{12}^{00}} & 0 & 0 & \mathcolor{blue}{S_{13}^{00}} \\ \mathcolor{blue}{S_{13}^{00}} & 0 & 0 & \mathcolor{magenta}{S_{12}^{00}}\\ 0 & \mathcolor{blue}{S_{13}^{00}} & \mathcolor{magenta}{S_{12}^{00}} & 0
    \end{bmatrix}\mathrm{,}\quad
    \bar{\bar{S}}^{\rm 10}=\begin{bmatrix} \mathcolor{purple}{S_{11}^{10}} & 0 &0 &\mathcolor{red}{S_{14}^{10}}\\0 & \mathcolor{yellow}{S_{22}^{10}} & \mathcolor{teal}{S_{23}^{10}} & 0 \\ 0 & \mathcolor{teal}{S_{23}^{10}} & \mathcolor{yellow}{S_{22}^{10}} & 0\\ \mathcolor{red}{S_{14}^{10}} & 0 & 0 & \mathcolor{purple}{S_{11}^{10}}
 \end{bmatrix},
 \label{Eq:sigmaz}
\end{equation}
and that the grating in the bottom row is symmetric under $C_{2y}$ leading to
\begin{equation}
    \bar{\bar{S}}^{\rm 00}=\begin{bmatrix} 0 & \mathcolor{magenta}{S_{12}^{00}} &\mathcolor{blue}{S_{13}^{00}} &\rm 0\\\mathcolor{magenta}{S_{12}^{00}} & 0 & 0 & \mathcolor{cyan}{S_{24}^{00}} \\ \mathcolor{blue}{S_{13}^{00}} & 0 & 0 & \mathcolor{magenta}{S_{12}^{00}}\\ 0 & \mathcolor{cyan}{S_{24}^{00}} & \mathcolor{magenta}{S_{12}^{00}} & 0
 \end{bmatrix}\mathrm{,}\quad 
    \bar{\bar{S}}^{\rm 10}=\begin{bmatrix} \mathcolor{purple}{S_{11}^{10}} & 0 &0 &\mathcolor{red}{S_{14}^{10}}\\0 & \mathcolor{yellow}{S_{22}^{10}} & \mathcolor{teal}{S_{23}^{10}} & 0 \\ 0 & \mathcolor{red}{S_{14}^{10}} & \mathcolor{purple}{S_{11}^{10}} & 0\\ \mathcolor{teal}{S_{23}^{10}} & 0 & 0 & \mathcolor{yellow}{S_{22}^{10}}
 \end{bmatrix}.
 \label{Eq:C2y}
\end{equation}
In these matrices, we have highlighted the coefficients with different colors to easily identify which terms are equal to each other. These colors also correspond to the colors of the arrows in Fig.~2 of the main text, where the matrices $\te{S}^{00}$ and $\te{S}^{10}$ correspond to the scattering responses depicted on left-hand side and right-hand side of the figure, respectively.

\section{Conical diffraction and propagation conditions}
\begin{figure}[h!]
\centering
\includegraphics[width=0.4\textwidth]{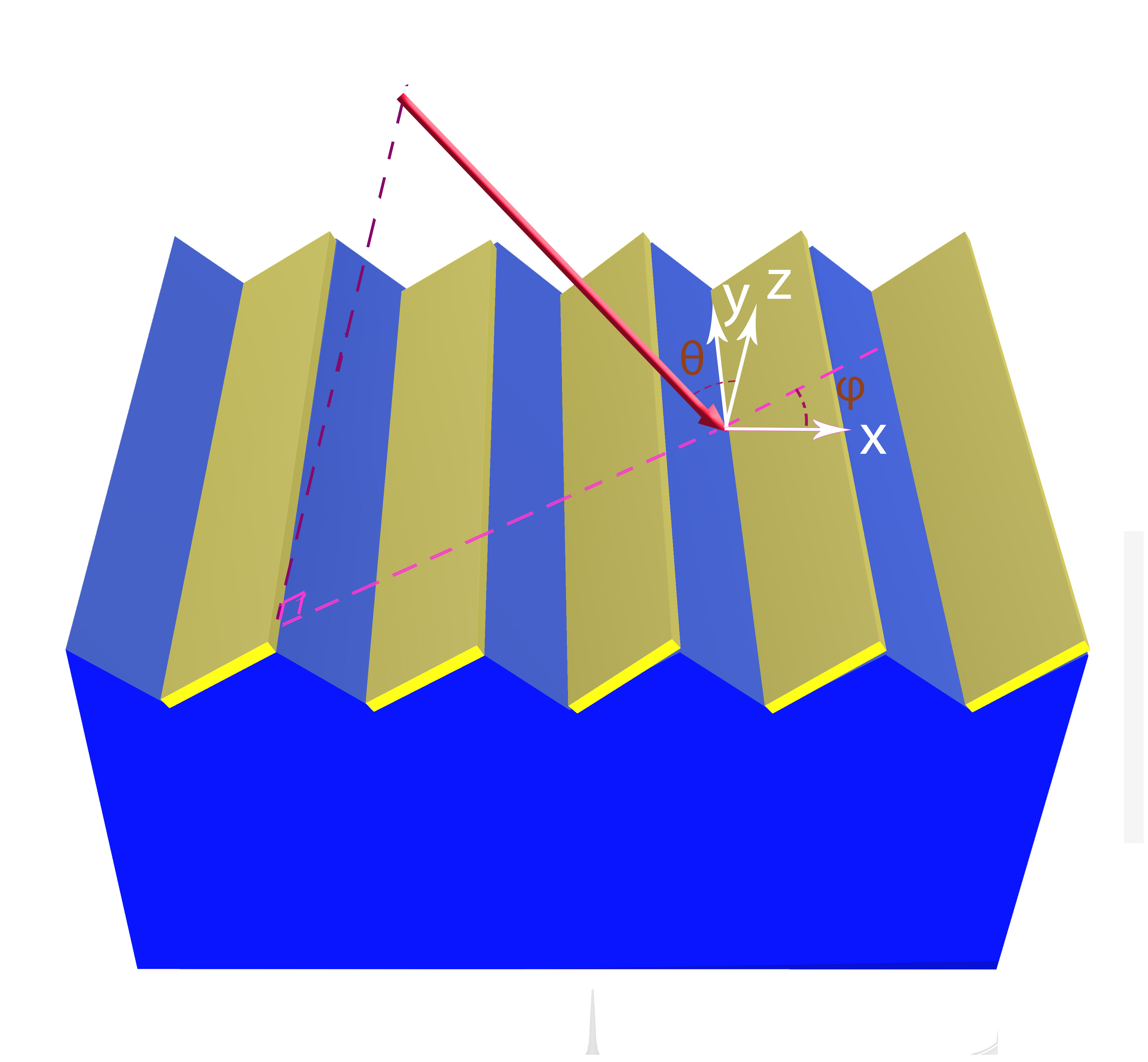}
\caption{ Three-dimensional schematic of the metagrating showing an arbitrary incidence plane.
}
\label{fig:3d_incidence}
\end{figure}

When the incidence plane is the XZ-plane as shown in Fig.\ref{fig:3d_incidence}, the grating equation both in the reflection and transmission regions is 
\begin{equation}
\label{eq_gratingEq}
     \Lambda (n_{\rm i} \sin(\theta_{\rm i})+n_{\rm out} \sin(\theta_{\rm m}))=-m\lambda_{\rm 0}
\end{equation}
where $\Lambda$, $ \theta_{\rm i}$, and $ \theta_{\rm m}$, $n_{\rm i}$, and $ n_{\rm out}$, $\Delta \phi$, and m are the grating period, the angle of incidence, the diffraction angle, the incidence medium's refractive index, the diffraction medium's refractive index, and an integer denoting the diffraction order, respectively.
\begin{equation}
 \sin(\theta_{\rm m})=\dfrac{\dfrac{-m\lambda_{\rm 0}}{\Lambda}-n_{\rm i}\sin(\theta_{\rm i})}{n_{\rm out}}
\end{equation}
Consequently, the cutoff condition for any diffraction mode to start propagation (i.e., $\theta_{\rm m}=90^{\circ}$) is:
\begin{equation}
 1=\dfrac{\dfrac{-m\lambda_{\rm 0}}{\Lambda}-n_{\rm i}\sin(\theta_{\rm i})}{n_{\rm out}}
\label{eq:cutoff}
\end{equation}
When the incidence plane is no longer restricted to the XZ-plane, the propagation conditions could be derived starting from symmetry considerations.

As it is depicted in Fig.\ref{fig:3d_incidence}, the wave vector for the incident plane wave is
\begin{equation}
\mathbf{k_{i}}=\alpha_{\rm i}\ \hat{\mathbf{x}}+\beta_{\rm i} \hat{\mathbf{y}}+\gamma_{\rm i} \hat{\mathbf{z}}=k_{0}n_{\rm i}(\cos(\phi_{\rm i})\sin(\theta_{\rm i}) \hat{\mathbf{x}}+\sin(\phi_{\rm i})\sin(\theta_{\rm i})\ \hat{\mathbf{y}}-\cos(\theta_{\rm i})\  \hat{\mathbf{z}})
\label{Eq:ki_3D}
\end{equation}
The wave vector of the diffracted mode is :
\begin{equation}
 \mathbf{k_{m}}=\alpha_{\rm m} \hat{\mathbf{x}}+\beta_{\rm m} \hat{\mathbf{y}}+\gamma_{\rm m} \hat{\mathbf{z}}
\end{equation}
From Bloch's theorem, the periodicity of the system  leads to the following :
\begin{equation}
 \mathbf{E}_{\rm inc}(x+a)=e^{i\alpha_{\rm i}\Lambda} \mathbf{E}_{\rm inc}(x)
\end{equation}
\begin{equation}
 \mathbf{E}_{\rm m}(x+a)=e^{i\alpha_{\rm m}\Lambda} \mathbf{E}_{\rm m}(x)
\end{equation}

The conservation of momentum law (i.e., the phase matching condition) and the boundary conditions of Maxwell's equations (i.e., the tangential electric field's continuity) leads to the following:
\begin{equation}
 \alpha_{\rm m}=\alpha_{\rm i}+\dfrac{2\pi m}{\Lambda}
\label{eq:alpham}
\end{equation}

The translation invariance along the y-axis implies :
\begin{equation}
 \beta_{\rm m}=\beta_{\rm i}
\label{eq:betam}
\end{equation}

The time translation symmetry (i.e., energy or frequency conservation) leads to the following:
\begin{equation}
     \gamma_{\rm m}=\pm \sqrt{k_{\rm 0}^{2}n_{\rm out}^{2}-\alpha_{\rm m}^2-\beta_{\rm m}^2}
\end{equation}

Hence, the propagation condition is

\begin{equation}
   |\gamma_{\rm m}|=|\mathbf{k_{\rm m}}\cdot \hat{\mathbf{n}}_{\rm gp}|=0
\label{Eq:cutoff_3D}
\end{equation}

where $\hat{\mathbf{n}}_{\rm gp}$ is the unit vector normal to the grating (i.e, the Z-axis).
We Substitute by Eq.\ref{Eq:ki_3D} into Eqs. \ref{eq:alpham},\ref{eq:betam}, and \ref{Eq:cutoff_3D}, we obtain 
\begin{equation}
   \cos(\phi_{\rm i})= \dfrac{k_{\rm 0}^2(n_{\rm out}^2-n_{\rm i}^2\sin^2(\theta_{\rm i}))-(\dfrac{2\pi m}{\Lambda})^2}{2k_{\rm 0}n_{\rm i}(\dfrac{2\pi m}{\Lambda})\sin(\theta_{\rm i})}
  \label{eq:cutoff_3d_phi}
\end{equation}
As a validation, it is worth noting that when $\phi=0$, Eq.\ref{eq:cutoff_3d_phi} is reduced to Eq.\ref{eq:cutoff}.

Also for  $ \overline{k_{x}}=\dfrac{\alpha_{\rm i}}{k_{\rm i}}=\dfrac{\alpha_{\rm i}}{k_{\rm 0}n_{\rm i}}$ and $ \overline{k_{y}}=\dfrac{\beta_{\rm i}}{k_{\rm i}}=\dfrac{\beta_{\rm i}}{k_{\rm 0}n_{\rm i}}$, we obtain
\begin{equation}
     \overline{k_{y}}^2=\dfrac{n_{\rm out}^2}{n_{\rm i}^2}-(\overline{k_{x}}^2+2\overline{k_{x}}\dfrac{m\lambda_{\rm 0}}{n_{\rm i}\Lambda}+(\dfrac{m\lambda_{\rm 0}}{n_{\rm i}\Lambda})^2)
    \label{eq:cutoff_moment}
\end{equation}

\section{Total transmittance for a perfect-electric-conductor strip}
\begin{figure}
\centering
\includegraphics[width=0.9\textwidth]{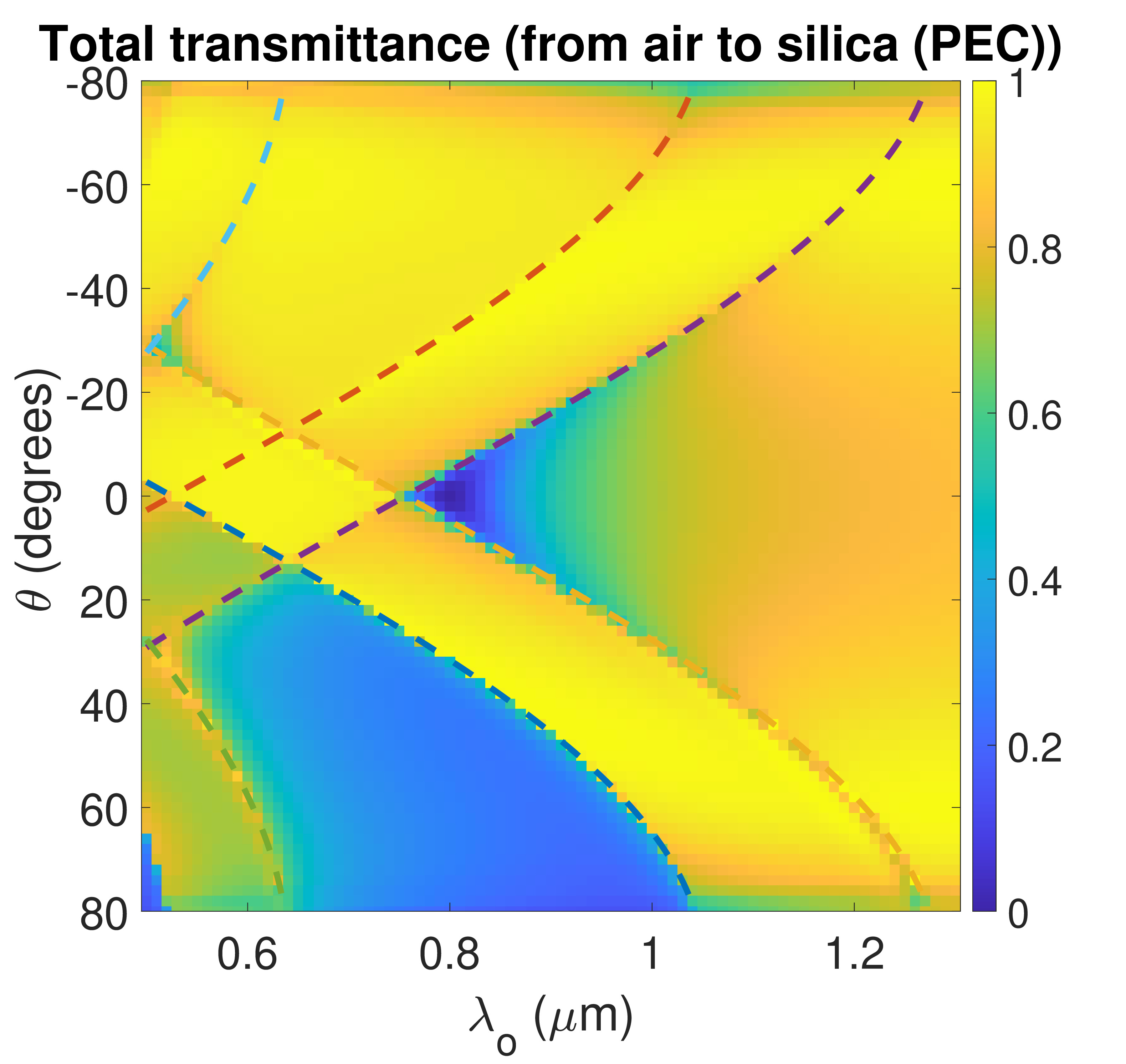}
\caption{Angular dispersion of the total transmittance for a strip with a perfect-electric-conductor boundary conditions.}
\label{fig:PIC}
\end{figure}

Figure~\ref{fig:PIC} shows the angular dispersion for the structure shown in Fig.4 (a) (in the main manuscript) with strips made of perfect electric conductor. Note that the asymmetry is weaker when compared to the gold material in the main manuscript. This shows the fact that the array of the gold strips is not simply an array of mirrors, but rather an array of tilted dipoles with their collective scattering response engineered to maximize the angular asymmetry.

\section{Zero-order transmittance and transmission phase}

\begin{figure}[!htb]
\centering
\begin{subfigure}[b]{0.485\textwidth}
         \centering
         \includegraphics[width=\textwidth]{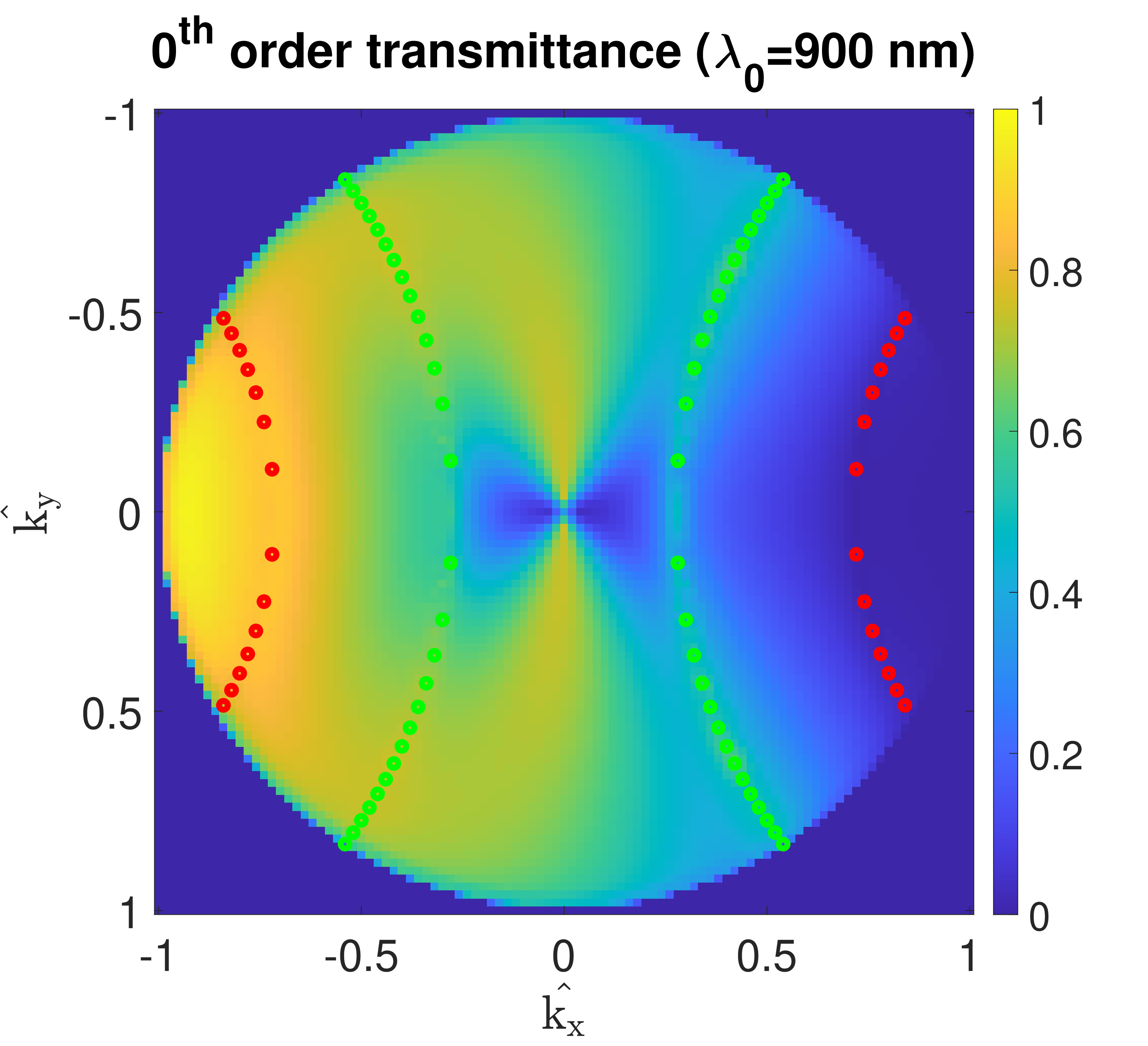}
         \caption{}
       
     \end{subfigure}
\begin{subfigure}[b]{0.48\textwidth}
         \centering
         \includegraphics[width=\textwidth]{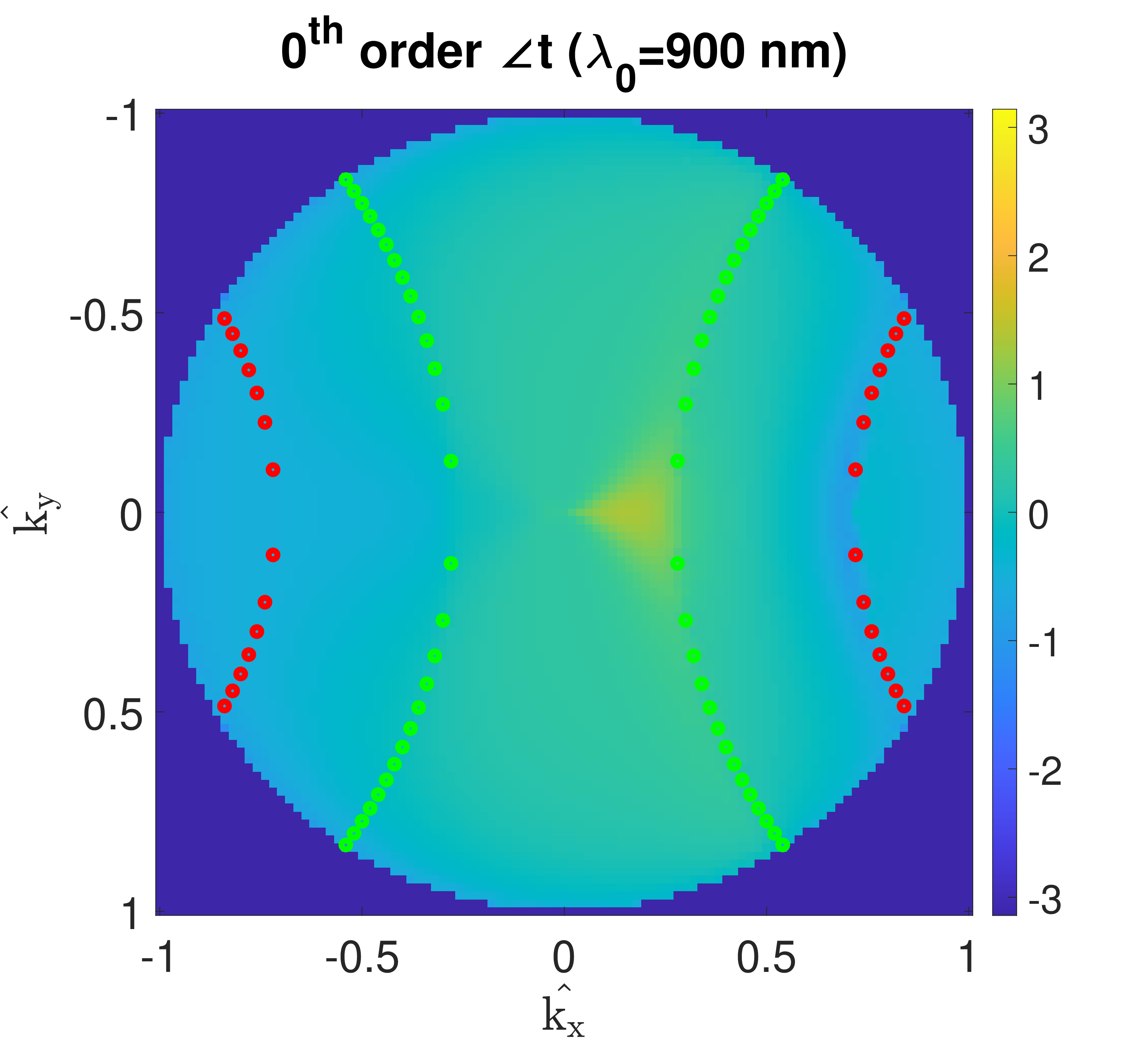}
         \caption{}
     
     \end{subfigure}

     \caption{(a) $\rm 0^{th}$ order transmittance, and (b) phase of the $\rm 0^{th}$ order transmission coefficient  for the study shown in Fig.5 (a) in the main manuscript.}
     \label{fig:zero_order_kxky}
\end{figure}

\begin{figure}
\centering
\includegraphics[width=0.9\textwidth]{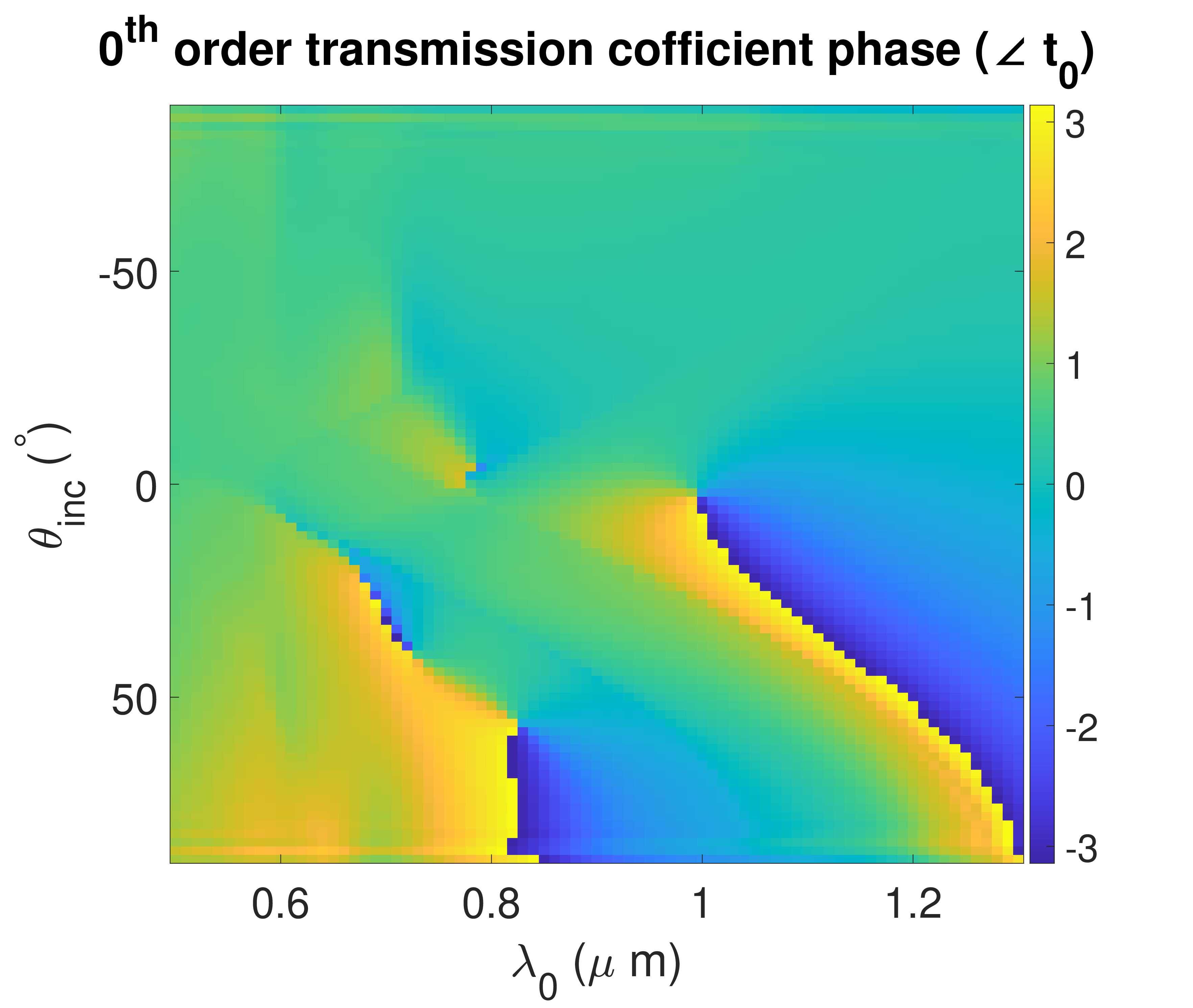}
\caption{Angular dispersion of the $\rm 0^{th}$ order transmission for the study shown in Fig.5 (a) in the main manuscript.}
\label{fig:zero_angular}
\end{figure}

Fig.\ref{fig:zero_order_kxky} (a) and (b) shown the $\rm 0^{th}$ order transmittance, and the $\rm 0^{th}$ order transmission phase, respectively for the study shown in Fig.5 (a) in the main manuscript. It can be seen that the total transmittance in Fig.5 (c) in the main manuscript is similar to Fig.\ref{fig:zero_order_kxky} (a), which implies that almost all of the energy goes to the $\rm 0^{th}$ order transmission channel. Fig.\ref{fig:zero_angular} shows the angular dispersion for the $\rm 0^{th}$ order transmission phase. It can be seen in Fig.\ref{fig:zero_order_kxky} (b) and Fig.\ref{fig:zero_angular} that the phase profile is uniform for the negative incidence angles (i.e, negative $\hat{k_{x}}$) where the structure is transmissive and transparent.

\section{Multipolar decomposition}
\begin{figure}
\centering
\includegraphics[width=0.9\textwidth]{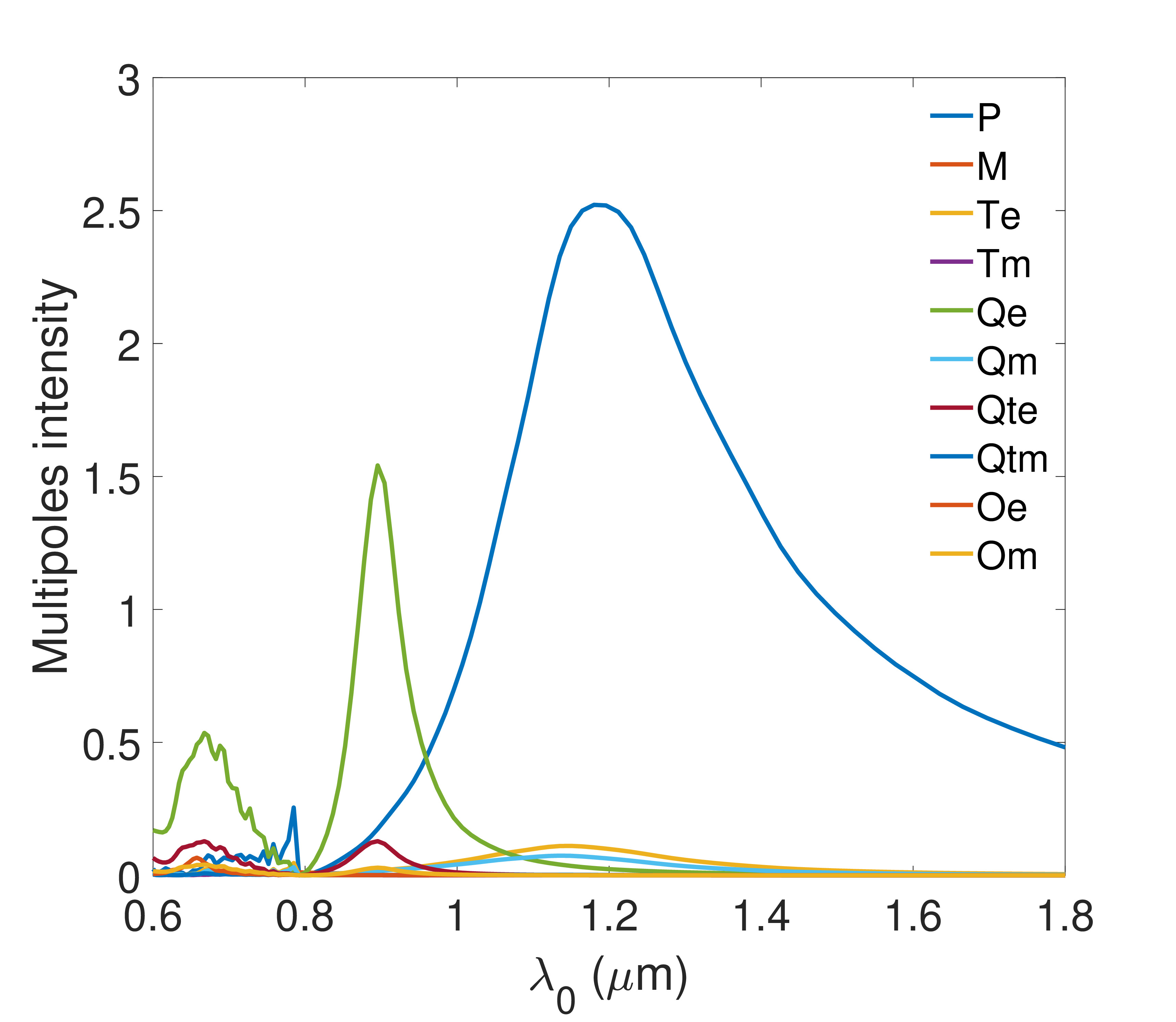}
\caption{Multipolar decomposition for the study in Fig.5 (e) in the main manuscript where the surrounding medium is silica.}
\label{fig:multipolar}
\end{figure}

Fig.\ref{fig:multipolar} shows the multipolar decompositon for the study in Fig.5 (e) in the main manuscript. It can be seen that the decomposition is almost an electrical dipole in the operation wavelength range.
\section{Process flow for the fabrication}

\begin{figure}
\centering
\includegraphics[width=0.9\textwidth]{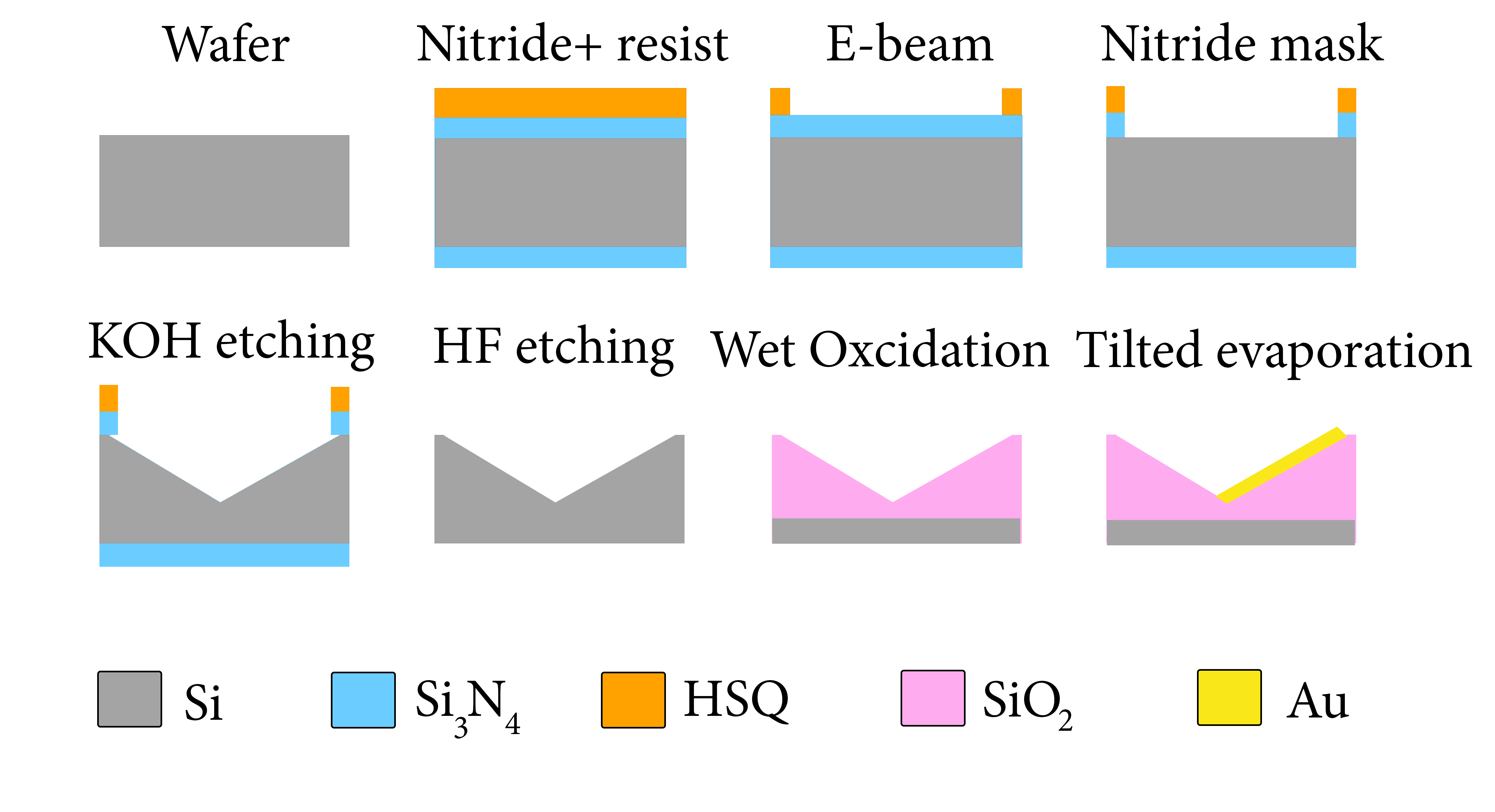}
\caption{Process flow for the fabrication}
\label{fig:process_flow}
\end{figure}

The process flow for the fabrication is shown in Fig.\ref{fig:process_flow}. First, we coat a silicon wafer with a nitride layer (thickness= 10 nm) on both sides. We then spin coat HSQ 2\% resist on the top side of the wafer with 2000 rpm to obtain a thickness of around 74 nm. A array of lines with a width of 60 nm and a period equal to 525 nm is then patterned over 1 $\rm cm^{2}$ by E-beam lithography. 
The sample is then developed in TMAH for two minutes. A nitride mask is created by plasma etching.
The sample is then etched in KOH 40\% 
to obtain an array of silicon triangles. This is owing to the fact that KOH etches silicon much faster in the 111 plane in comparison to the 100 plane, while it does not etch silicon nitride. 

Then, the sample is immersed in HF 50\% 
to remove the nitride layer on both sides. Thermal wet oxidation is then used to oxidize the wafer for a depth of 1.5 $\rm \mu m$ (i.e., silicon becomes silica) while perserving the geometrical pattern of the triangular ridges. A dedicated holder is then used to fix the sample at an oblique angle inside the gold evaporation machine. 
We evaporate a 2 nm layer of chromium as an adhesive layer for the gold on the sample. Finally, we evaporate a gold layer on only side of the triangular silica strips (with a thickness of 22 nm). 

\section{Optical setup and experimental measurements}
\label{sec:exp}
\begin{figure}
\centering
\includegraphics[width=0.8\textwidth]{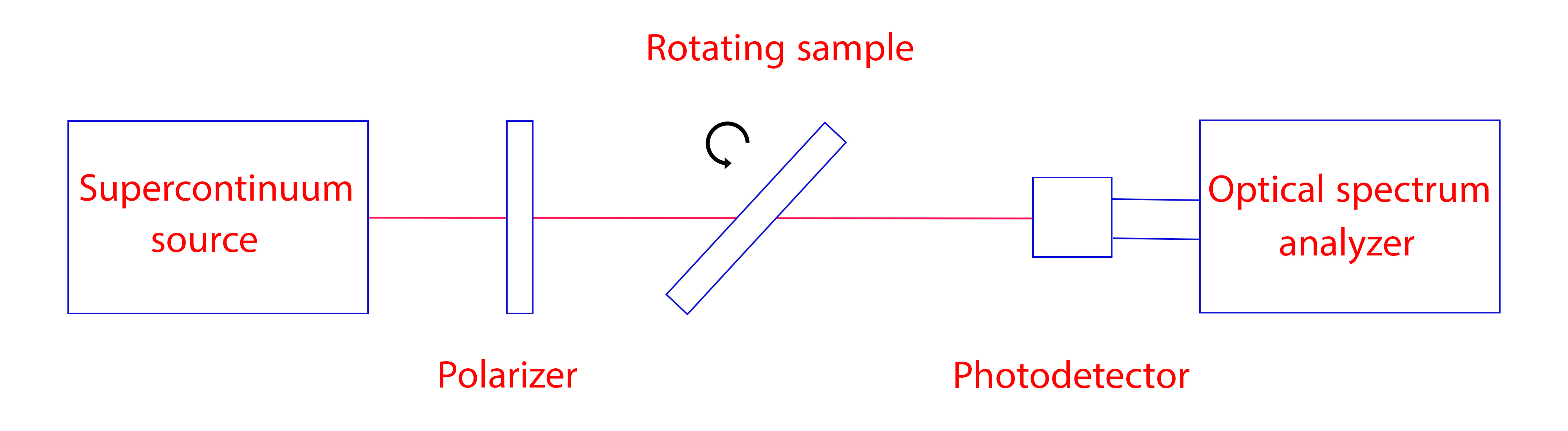}
\caption{Top view of the optical setup used for measurement.}
\label{fig:setup}
\end{figure}

The optical setup used for getting the measurements of the transmittance as a function of the incidence angle and the operation wavelength is made of a super continuum source (SuperK Extreme EXW-12), a polarizer, a rotating wafer holder, and an optical spectrum analyzer (YOKOGAWA AQ6370D). As depicted in Fig.\ref{fig:setup}, the zeroth-order transmittance could be easily measured using the optical spectrum analyzer since the zeroth-order transmittance must be always parallel to the beam axis after going through the wafer.

\section{Fabrication characterization}
\begin{figure}
\centering
\includegraphics[width=0.7\textwidth]{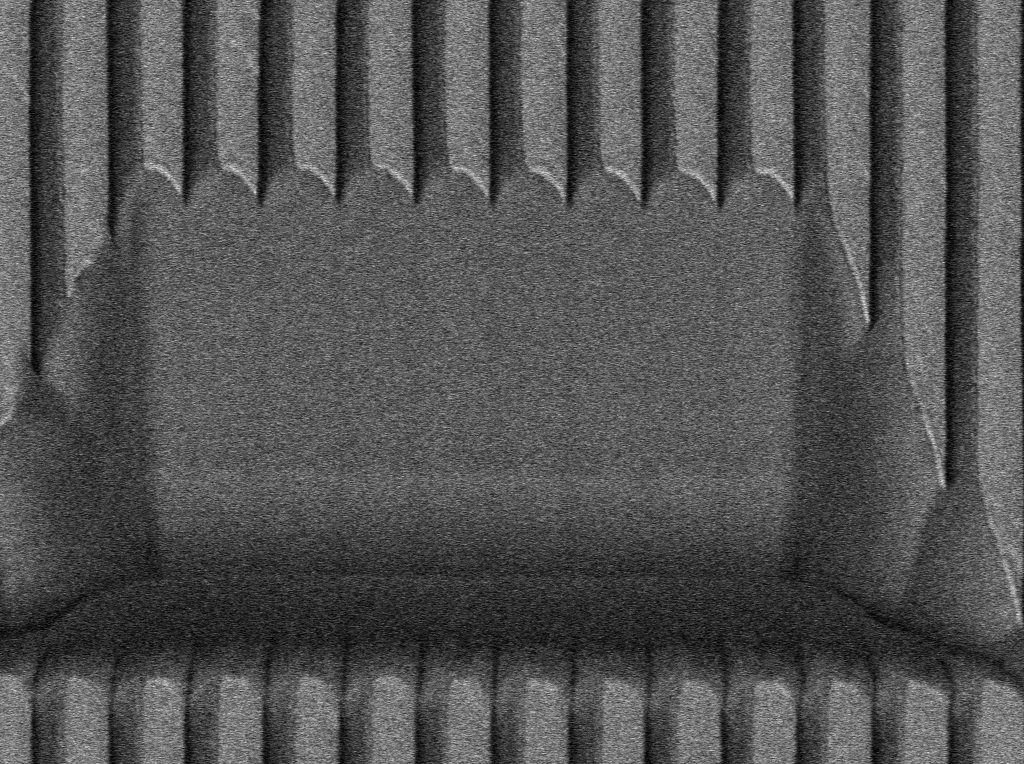}
\caption{A FIB cut of the sample showing the curvature of the strips after oxidation.}
\label{fig:char}
\end{figure}

\begin{figure}[!htb]
\centering
\begin{subfigure}[b]{0.32\textwidth}
         \centering
         \includegraphics[width=\textwidth]{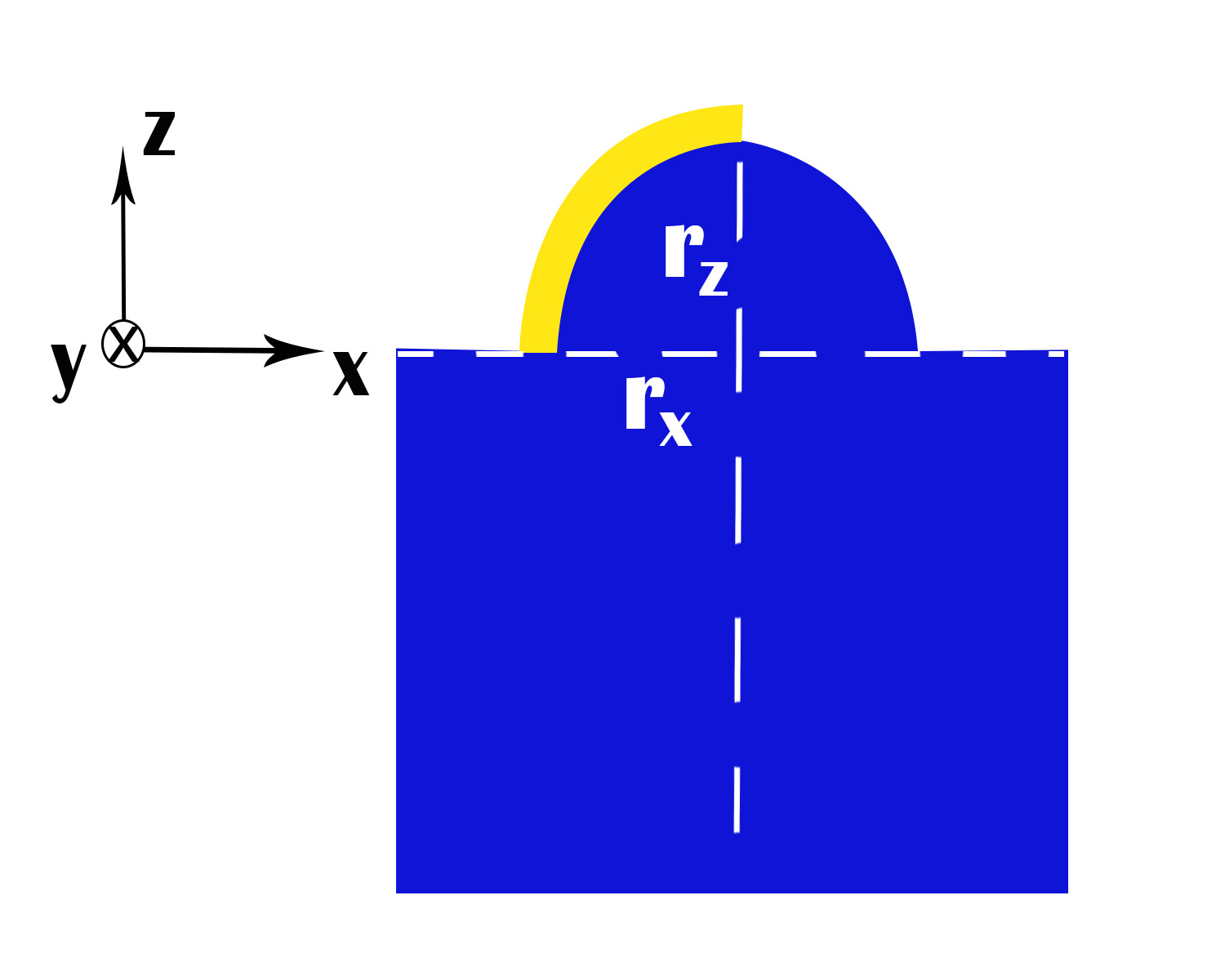}
         \caption{}
         \label{fig:N0_band_theo}
     \end{subfigure}
\begin{subfigure}[b]{0.32\textwidth}
         \includegraphics[width=\textwidth]{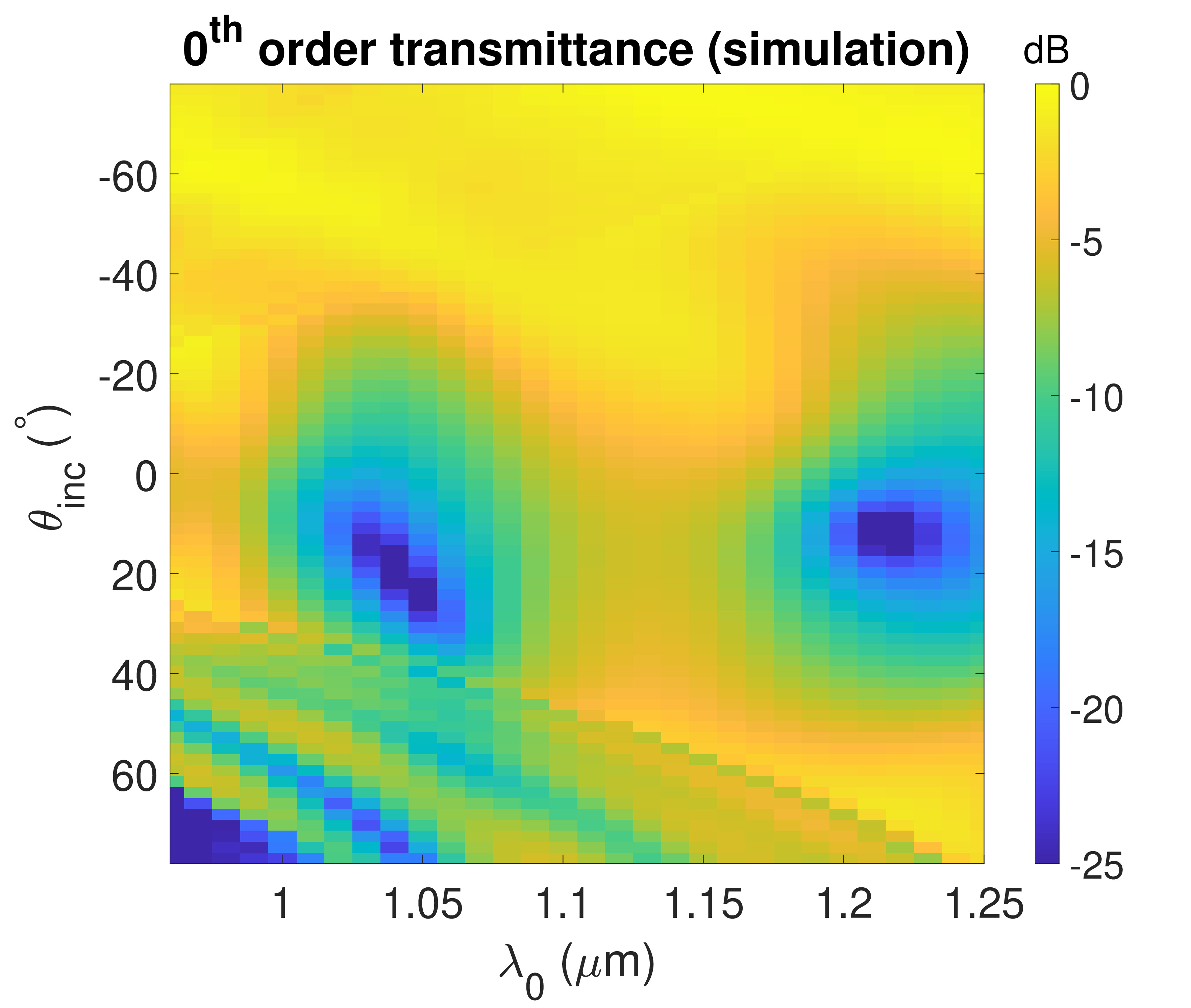}
         \caption{}
         \label{fig:N0_band_exp}
     \end{subfigure} 
     
\begin{subfigure}[b]{0.32\textwidth}
         \centering
         \includegraphics[width=\textwidth]{zero_order_transmittance_angle_sim.jpg}
         \caption{}
         \label{fig:N0_band_theo}
     \end{subfigure}
\begin{subfigure}[b]{0.32\textwidth}
         \includegraphics[width=\textwidth]{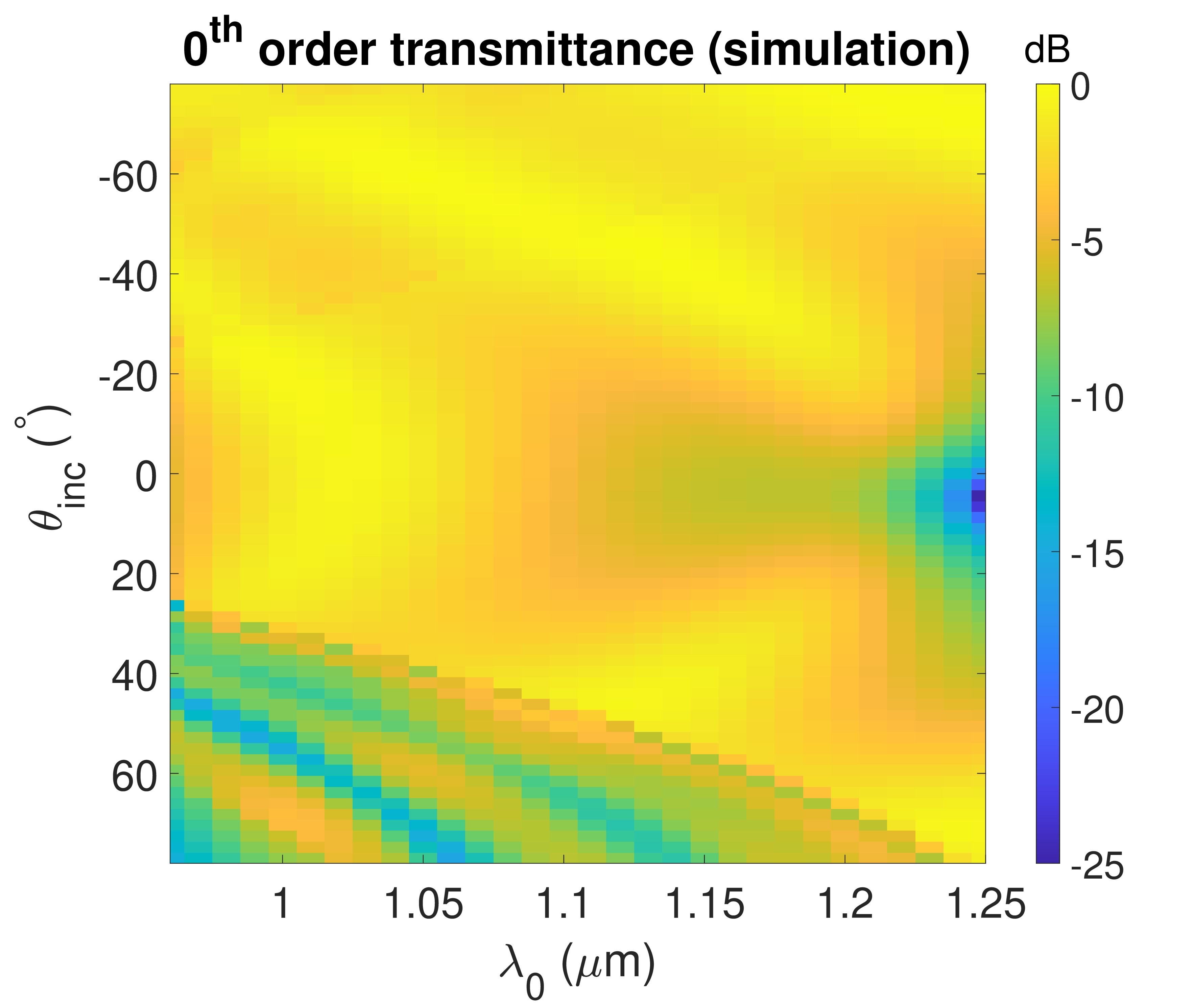}
         \caption{}
         \label{fig:N0_band_exp}
     \end{subfigure}

\centering

     \caption{(a) Schematic showing the radii in the X- and Z- directions (i.e., $r_{x}$ and $r_{z}$). (b), (c), and (d) The simulated angular dispersion of the zeroth-order transmittance for a curved strip with (b) $r_{x}=\Lambda/2$ and $r_{z}=\Lambda/2$, (c) $r_{x}=\Lambda/2$ and $r_{z}=1.3\Lambda/2$, and (d) $r_{x}=0.9\Lambda/2$ and $r_{z}=1.4\Lambda/2$. The oxidation thickness below the curved strips is equal to 1800 nm.}
     \label{Fig:curvature_sim}
\end{figure}

Here, we explain the origin of the difference between the experimental measurements and the results obtained by full-wave simulations. As depicted in Fig.\ref{fig:char}, the tilted strips have some curvature due to the thermal oxidation step. The angular dispersion of the absorbance in the non-diffractive regime, and the transmittance as a result, are very sensitive to the curvature of the tilted strips as depicted in Fig.\ref{Fig:curvature_sim} for three simulated cases. Consequently, the simulated angular dispersion of the transmittance deviates from the measured data in the non-diffractive regime owing to this sensitivity.

